\documentclass[times,final]{./tex-library/elsarticle}
\usepackage[margin=1in]{geometry}

\usepackage{framed,multirow}

\usepackage{amssymb}
\usepackage{amsmath}
\usepackage{latexsym}
\usepackage{booktabs}

\usepackage{multirow}
\usepackage{mathtools}
\usepackage{graphicx}
\usepackage{colortbl}
\usepackage{empheq}
\usepackage{bm}
\usepackage{wasysym}
\usepackage[usenames,dvipsnames]{xcolor}
\usepackage{caption,subcaption}

\usepackage[toc,page]{appendix}
\usepackage[colorlinks]{hyperref}

\usepackage{./tex-library/dsfont}
\usepackage[capitalise]{./tex-library/cleveref}

\crefname{appsec}{Appendix}{Appendices}

\usepackage{url}
\usepackage{xcolor}
\definecolor{newcolor}{rgb}{.8,.349,.1}

\newcommand{\ReN}{\mbox{\textit{Re}}}
\newcommand{\PeN}{\mbox{\textit{Pe}}}
\newcommand{\RaN}{\mbox{\textit{Ra}}}
\newcommand{\GrN}{\mbox{\textit{Gr}}}
\newcommand{\PrN}{\mbox{\textit{Pr}}}
\newcommand{\ScN}{\mbox{\textit{Sc}}}
\newcommand{\SC}{\mbox{\textit{SC}}}
\newcommand{\GSC}{\mbox{\textit{GSC}}}
\newcommand{\grad}{\nabla}
\DeclarePairedDelimiter\floor{\lfloor}{\rfloor}

\newcommand{\first}{$1^\textnormal{st}$}
\newcommand{\second}{$2^\textnormal{nd}$}
\newcommand{\third}{$3^\textnormal{rd}$}
\newcommand{\fourth}{$4^\textnormal{th}$}
\newcommand{\tR}[1]{{#1}}
\newcommand{\tO}[1]{{#1}}

\begin{document}


\begin{frontmatter}

\title{A stable and accurate scheme for solving the Stefan problem coupled with natural convection using the Immersed Boundary Smooth Extension method}%

\author[1,3]{Jinzi Mac Huang\corref{cor1}}\cortext[cor1]{Corresponding author. Email: machuang@nyu.edu}
\author[1,2]{Michael J. Shelley}
\author[2]{David B. Stein}

\address[1]{Applied Mathematics Laboratory, Courant Institute, New York University, New York, NY 10012, USA}
\address[2]{Center for Computational Biology, Flatiron Institute, Simons Foundation, New York, NY 10010, USA}
\address[3]{NYU-ECNU Institute of Physics and Institute of Mathematical Sciences, New York University Shanghai, Shanghai, 200122, China}


\begin{abstract}
The dissolution of solids has created spectacular geomorphologies ranging from centimeter-scale cave scallops to the kilometer-scale ``stone forests'' of China and Madagascar. Mathematically, dissolution processes are modeled by a Stefan problem, which describes how the motion of a phase-separating interface depends on local concentration gradients, coupled to a fluid flow. Simulating these problems is challenging, requiring the evolution of a free interface whose motion depends on the normal derivatives of an external field in an ever-changing domain. Moreover, density differences created in the fluid domain induce self-generated convecting flows that further complicate the numerical study of dissolution processes. In this contribution, we present a numerical method for the simulation of the Stefan problem coupled to a fluid flow. The scheme uses the Immersed Boundary Smooth Extension method to solve the bulk advection-diffusion and fluid equations in the complex, evolving geometry, coupled to a $\theta$-$L$ scheme that provides stable evolution of the boundary. We demonstrate \third-order temporal and pointwise spatial convergence of the scheme for the classical Stefan problem, and \second-order temporal and pointwise spatial convergence when coupled to flow. Examples of dissolution of solids that result in high-Rayleigh number convection are numerically studied, and qualitatively reproduce the complex morphologies observed in recent experiments.
\end{abstract}


\end{frontmatter}



\section{Introduction}

Mass exchange between material phases, as in melting, solidification, and dissolution, drives the evolution of the phase separating interface. When the driving dynamics are generated by diffusion or heat transfer, such interface-evolution problems are categorized as Stefan problems \cite{rubinvsteuin2000stefan,meakin2010geological, moore2017riemann}. In the cases of solidification and melting, heat flux at the interface drives the boundary motion \cite{moore2017riemann, vanier1970free, favier_purseed_duchemin_2019}; in the case of dissolution, molecular diffusion converts the solid into solute \cite{moore2017riemann,FLM:9533822, wykes2018self, claudin_duran_andreotti_2017, ristroph_2018}. The rate at which melting or dissolution occurs depends on the distribution of heat or concentration in the fluid phase, and this distribution is typically governed by an advection-diffusion equation.

These processes cannot be fully described by the classical Stefan problem alone: diffusion of heat/solute induces density changes in the fluid, causing buoyancy driven flows that reorganize the temperature/concentration fields \cite{childress2009introduction, tritton2012physical}. These rearrangements induce Rayleigh-Taylor \cite{boffetta2017incompressible} and Rayleigh-B\'{e}nard convection \cite{ahlers2009heat} and hence enhance the interfacial melting/dissolving rate \cite{favier_purseed_duchemin_2019, wykes2018self,ristroph_2018}, further complicating the interfacial dynamics. These complex dynamics have been shown to generate fascinating pattern formations such as the scalloping of icebergs \cite{gilpin1980wave}, the roughening of dissolving surfaces \cite{wykes2018self,claudin_duran_andreotti_2017,ristroph_2018}, and the sharpening of dissolving pinnacles \cite{Huang23339}.

A significant body of literature exists for the numerical solution of Stefan problems, including schemes based on the \tR{level-set method \cite{osher2001level, gibou2003level, CHEN19978, javierre2006comparison,Gibou2005577,chen2009numerical}, phase field method \cite{favier_purseed_duchemin_2019, javierre2006comparison}, and others \cite{javierre2006comparison, beckett2001moving}.} In this paper, we seek to construct a scheme for solving the Stefan problem with convection in a high Reynolds number fluid. While difficulties are numerous, the two primary ones stem from the fact that (1) it has been observed that these problems typically generate flows and interfacial boundaries with fine length-scales; (2) the boundary evolution depends on gradients of computed quantities at the boundary \cite{wykes2018self,claudin_duran_andreotti_2017,ristroph_2018}. The first difficulty demands a high resolution scheme that can accommodate a moving boundary, and thus a natural choice is to use an embedded boundary method. These schemes embed the complex and evolving boundary in a larger, geometrically simple domain, enabling the use of many of the fast and robust methods for solving partial differential equations (PDE) that have been developed for regular domains. Unfortunately, most embedded boundary schemes, such as the Immersed Boundary Method \cite{peskin2002immersed}, fail to accurately capture the gradient of unknowns at the boundary. For the Immersed Boundary Method as applied to the diffusion equation, for example, normal derivatives of the concentration field are inconsistent at the boundary \cite{Stein2016252}. The use of such a method would thus yield inaccurate dynamics of the phase interface.

To construct an accurate, fast, and stable solver, we leverage the recent development of the Immersed Boundary Smooth Extension (IBSE) method \cite{Stein2016252,Stein2017155,STEIN201956}, an embedded boundary scheme that accurately captures derivative information at the boundary. The bulk advection-diffusion equations for heat/solute transport and the Navier-Stokes equations are solved using the IBSE method in the evolving fluid region. The interfacial boundary dynamics is solved using the $\theta$-$L$ method \cite{hou1994removing,alben2002drag,moore2013self} and the Gibbs-Thomson effect is added to stabilize its long-time evolution. The solver is accurate (at least second-order pointwise in space), efficient and stable -- allowing for high resolution of flow and boundary features, as well as long-time simulation of complex phenomena. We validate the solver through comparisons to analytic solutions and refinement studies, demonstrate that it reproduces classical instabilities, and use it to analyze the formation of complex surface morphologies and fluid flows for a dissolving solid.

This paper is organized as follows. In \cref{stefan_with_natural_convection} we define the Stefan problem for melting and dissolution, and discuss its coupling to an external fluid. In \cref{methods}, we describe the IBSE-Stefan solver; including a review of the IBSE method introduced in \citep{Stein2016252,Stein2017155}, as well as the $\theta$-$L$ method, introduced in \cite{hou1994removing,alben2002drag,moore2013self}. In \cref{numImp}, we describe the numerical method and discuss important details of the specific implementation. In \cref{stefanSection}, we compare the numerical solution of a Stefan problem without convection to an analytic solution and show that the solver captures the Mullins-Sekerka instability, resolving complex dynamics of the solid interface. In \cref{dissolution_results}, we show several examples of dissolution coupled with Navier-Stokes flows. Sweeping across a range of Reynolds and P\'{e}clet numbers, we explore which control parameters are most predictive of pattern formation on the interface and identify that the Rayleigh number governs the regimes of shape dynamics. In \cref{discussion}, we discuss the scope and limitations of our numerical method, and demonstrate that our study qualitatively reproduces phenomena seen in recent experiments \cite{wykes2018self,claudin_duran_andreotti_2017}.

\section{Stefan problem coupled with natural convection}
\label{stefan_with_natural_convection}

\subsection{Stefan problem}
\label{intro2Stefan}
\begin{figure}
	\centering
	\includegraphics[width=0.6\textwidth]{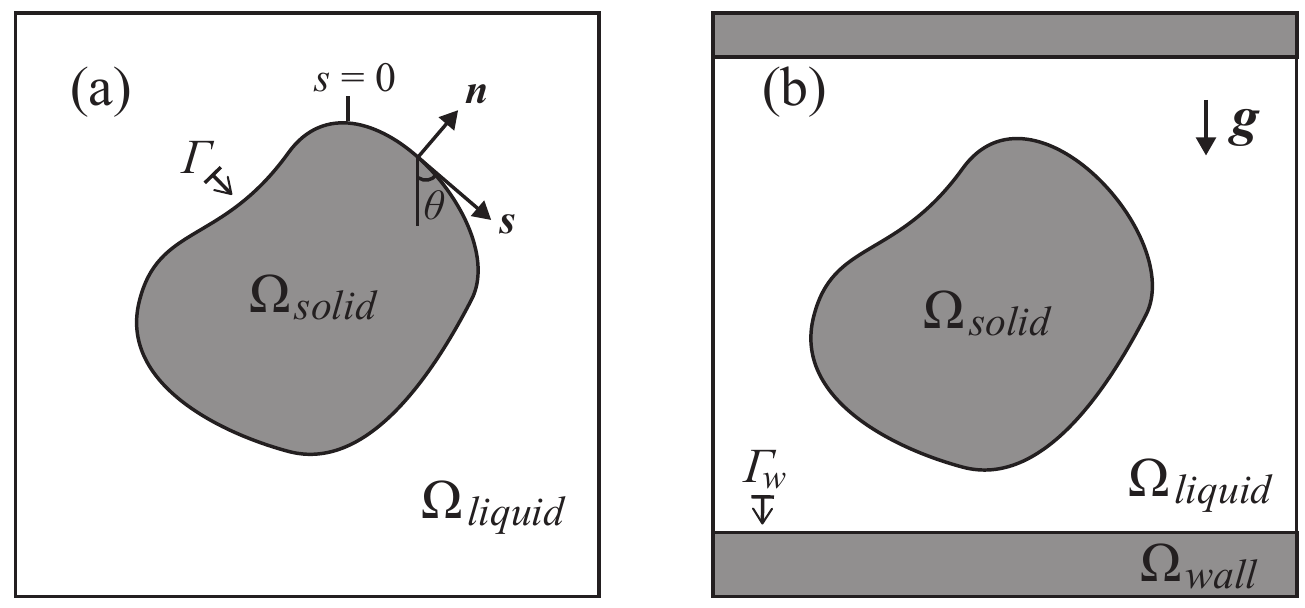}
	\caption{Schematic for the periodic computational domain $\mathbb{T}^2 = [0, 2\pi]\times[0,2\pi]$. The solid phase is shown in gray and the liquid phase is shown in white. (a) For the classical Stefan problem without the presence of flow, the solid domain $\Omega_{solid}$ and liquid domain $\Omega_{liquid}$ are separated by the evolving interface $\Gamma(t)$. (b) For the Stefan problem with flow, the gravity $\mathbf{g}$ induces convection. Rigid walls $\Omega_{wall}$ are added to bound the liquid domain from above and below.}
	\label{fig1}
\end{figure}

The classical Stefan problem models the diffusion of heat between two phases of a substance that are separated by a phase interface $\Gamma$, along with the evolution of that interface. Physically, one phase corresponds to a solid domain $\Omega_{solid}$ as shown in \cref{fig1} that solidifies from or melts into its liquid phase, depending on the direction of heat flux. In the liquid phase $\Omega_{liquid}$, the heat transfer is modeled by the heat equation
\begin{equation}
\label{heateq}
\frac{\partial T}{\partial t} = \nabla\cdot ( K \nabla T ) \quad \mbox{ in  } \Omega_{liquid}. 
\end{equation}
Here $T(\mathbf{x}, t)$ is the temperature field and $K$ is the thermal diffusivity of the liquid phase. The solid phase is typically assumed to have infinite thermal conductivity so that
\begin{equation}
	T = T_{m} \quad \mbox{ in  } \bar{\Omega}_{solid}, 	
\end{equation}
where $T_m$ is the melting point of the solid. This condition provides a boundary condition for \cref{heateq}: $T = T_m$ on $\Gamma \subset \bar{\Omega}_{solid}$. 

When the liquid temperature is greater than $T_m$ the solid melts into the liquid and the boundary recedes. If the liquid temperature is lower than $T_m$, which means the liquid is in an undercooled state, solidification occurs at the boundary and the solid grows. This motion is a consequence of Fourier's law of heat transfer, which specifies that the rate of material transfer is proportional to the heat flux $\mathbf f = - K \rho c_p \nabla T $, with $\rho$ and $c_p$ the liquid density and specific heat, respectively. Combining Fourier's law with the conservation of mass leads to a normal boundary velocity  $V_n(s) = - (K c_p / l)  \partial T/\partial n$ at the interface, where $l$ is the latent heat of the liquid. 

In the following, we assume that the solid and liquid are defined in the periodic domain $\mathbb{T}^2 = [0, 2\pi]\times[0,2\pi]$ and the solid domain $\Omega_{solid}$ is simply connected, as shown in \cref{fig1}(a). The moving interface $\Gamma(t)$ between the solid and liquid phases is assumed to be smooth, and parameterized by the arclength parameter $s\in[0,L)$, where $L(t)$ gives the circumference of $\Gamma$. Rescaling time as $t' = t / K$, temperature as $c = (T - T_m) / \Delta T$ (where $\Delta T = \underset{t=0}{\max}(|T - T_{m}|)$), and dropping the $'$ notation on $t$ gives the dimensionless equations
\begin{subequations}
\label{stefan}
\begin{align}
     \frac{\partial c}{\partial t} = \Delta c &\quad \mbox{ in  } \Omega_{liquid},\\
     c = c_m &\quad \mbox{ on  } \Gamma(t),\\
     V_n =  \beta \frac{\partial c}{\partial n} &\quad \mbox{ on  } \Gamma(t), \\
     c(\mathbf{x}, 0) = c_0  &\quad \mbox{ in  } \Omega_{liquid}.
\end{align} 
\end{subequations}

It may appear that there is one too many boundary conditions for $c$. Note however that one serves to fix $c$, while the other defines the evolution of the time-dependent interface $\Gamma$. Only one parameter -- the Stefan number $\beta = - c_p \Delta T/l$ -- controls the dynamics of the dimensionless system.  For the solidification problem ($c_0 \in [-1,0], c_m = 0 \mbox{ on } \Gamma$), it has been observed that the Stefan problem posed in \cref{stefan} undergoes a Mullins-Sekerka instability, caused by rapid growth of the interface in high-curvature regions. This phenomena will be further demonstrated with numerical examples in \cref{stefanSection}. 

It is worth noting that \cref{stefan} also models the dissolution process, where the temperature field is replaced with the concentration field $c\in [0,1]$. The parameter space in melting, solidification and dissolution is listed below.

\begin{center}
\begin{tabular}{ | c | c| c | c| } 
\hline
Process & $\beta$ & range of $c_0$ & $c_m$ \\ 
\hline
melting & $-c_p \Delta T/l$ & $[-1,0]$ & 0\\ 
\hline
solidification & $-c_p \Delta T/l$& $[0,1]$ & 0\\ 
\hline
dissolution & $\rho/\rho_s$ & $[0,1]$ & 1\\ 
\hline
\end{tabular}
\end{center}

The parameter $\rho_s$ in the case of dissolution is the solid density. The Stefan number $\beta$ is positive for dissolution processes as lower liquid concentration $c_0$ leads to dissolution hence $V_n < 0$; however $\beta<0$ for the melting and solidification processes since lower liquid temperature leads to solidification and $V_n > 0$. In the following sections we will neglect to distinguish between the heat and the mass transfer problems, instead referring to them simply as Stefan problems.

\subsection{The Stefan problem with natural convection}

In the liquid phase, temperature or solute concentration differences lead to changes in the specific volume of fluid parcels, and hence the fluid density. These changes in density lead to the buoyancy driven convection that typically accompanies melting and solidification processes. Using the Boussinesq approximation \cite{tritton2012physical, peyret2013spectral}, we treat the fluid as incompressible and subject to a \tR{buoyancy force $\mathbf{B} \propto - c\hat{\mathbf{y}}$} that is proportional to the concentration difference. The concentration field $c(\mathbf{x}, t)$ diffuses and is advected by the flow field $\mathbf{u}(\mathbf{x}, t)$. The complete model in dimensionless form is given by
\begin{subequations}
\label{stefan-nc}
\begin{align}
     \frac{\partial c}{\partial t} + \mathbf{u}\cdot\nabla c = \frac{1}{\PeN}\Delta c &\quad \mbox{ in  } \Omega_{liquid},\\
     \frac{\partial \mathbf{u}}{\partial t} + \mathbf{u}\cdot\nabla \mathbf{u} =\frac{1}{\ReN} \Delta \mathbf{u} - \nabla p - c \hat{\mathbf{y}} &\quad \mbox{ in  } \Omega_{liquid}, \label{NSE-approx}\\\
     \nabla \cdot \mathbf{u} = 0 &\quad \mbox{ in  } \Omega_{liquid},\\
     V_n =  \frac{\beta}{\PeN}\frac{\partial c}{\partial n} &\quad \mbox{ on  } \Gamma, \\
     \mathbf{u} = 0,\ c = 1 &\quad \mbox{ on  } \Gamma, \label{bc_on_surface}\\
     \mathbf{u}(\mathbf{x}, 0) = \mathbf{u}_0(\mathbf{x}), c(\mathbf{x}, 0) = c_0(\mathbf{x})  &\quad \mbox{ in  } \Omega_{liquid}.
\end{align} 
\end{subequations}

In the experiments of dissolution that motivated this numerical study \cite{wykes2018self}, dissolution occurred in a tank, with solid boundaries at the bottom (and an air-liquid interface at the top). To approximate these conditions, we place adiabatic walls $\Gamma_{wall}$ at the top and bottom of the domain so that the liquid is bounded from above and below \tR{but periodic in the horizontal direction; see \cref{fig1}(b). At these walls, fluid motion ceases and the walls are impermeable to the solute, i.e. we impose that $\mathbf{u} = 0$ and $\partial_n c = 0$ on $\Gamma_{wall}$.} The behavior of this system is controlled by the Reynolds number ($\ReN$) of the fluid and the P\'{e}clet number ($\PeN$) of the advection diffusion equation, and the parameter $\beta$ that governs the rapidity of boundary evolution. \tR{Because time is rescaled by the typical flow speed, the diffusivity is now $1/\PeN$ instead of $1$ in \cref{stefan}, and the Stefan number in this problem is $\beta / \PeN$ instead of $\beta$.} These parameters are typically related to the Grashof number $\GrN$ and Prandtl (or Schmidt) number $\PrN$ (or $\ScN$) as $\ReN = \sqrt{\GrN}$, $\PeN = \sqrt{\GrN} \PrN$ (or $\sqrt{\GrN} \ScN$) \cite{schlichting2016boundary}. In heat transfer problems, $ \GrN = \alpha_V g L^3 \Delta T / \nu^2$ and $ \PrN = \nu/K$, with $\alpha_V$, $g$ and $\nu$ the thermal expansion coefficient, acceleration due to gravity and kinematic viscosity respectively. In mass transfer problems, $ \GrN = \beta_V g L^3 / \nu^2$ and $ \ScN = \nu/D$, with $\beta_V$, $g$, $D$ and $\nu$ the relative density difference between the solid and the fluid, acceleration due to gravity, molecular diffusivity and kinematic viscosity respectively. As we will show later in \cref{dissolution_results}, an important control parameter of dissolution under natural convection is the Rayleigh number $\RaN = \GrN\ScN = \ReN\PeN = \beta_V g L^3 / \nu D$, which governs the regimes of dissolution shape dynamics. 

For the familiar case of ice melting into water \cite{vanier1970free} at room temperature, $\GrN$ is in the range of $10^6$ - $10^7$, $\PrN$ is around 1 and $\beta$ is around $0.1$, leading to a Reynolds number $\ReN \sim 10^3$, a P\'{e}clet number $\PeN \sim 10^3$ and a Rayleigh number $\RaN\sim 10^6$. For candy dissolving into water \cite{wykes2018self}, $\GrN$ is in the range of $10^8$, $\ScN$ is around $10^3$, and $\beta$ is around $1$, leading to a Reynolds number $\ReN \sim 10^4$, a P\'{e}clet number $\PeN \sim 10^7$ and a Rayleigh number $\RaN\sim 10^{11}$. The following table summarizes the definition and typical ranges of the parameters for these cases.

\begin{center}
\begin{tabular}{ | c | c| c | c| c| c|} 
\hline
Process & $\GrN$ & $\PrN$ or $\ScN$ & $\ReN$ & $\PeN$ & $\RaN$ \\ 
\hline
melting or solidification & $ \alpha_V g L^3 \Delta T / \nu^2$ & $\PrN = \nu/K$ & $ \sqrt{\GrN}$ & $\sqrt{\GrN} \PrN$ & $\GrN\PrN$ \\ 
typical range \cite{vanier1970free} &  $10^6-10^7$ & $1-10$ & $10^3-10^4$ & $10^3-10^4$& $10^6-10^8$ \\ 
\hline
dissolution of sugar into water  & $\beta_V g L^3 / \nu^2$ & $ \ScN = \nu/D$  & $ \sqrt{\GrN}$ & $\sqrt{\GrN} \ScN$ & $\GrN\ScN$\\ 
typical range \cite{wykes2018self} & $10^8-10^{10}$ & $10^3-10^4$ & $10^4-10^5$ & $10^7-10^9$ & $10^{11}-10^{14}$\\ 
\hline
\end{tabular}
\end{center}

We remark that the $\mathbf{u}=0$ boundary condition given in \cref{bc_on_surface} is not exactly a no-slip condition because the interface is moving. However, in all simulations in this paper and most dissolution processes in nature, the Stefan number $\beta / \PeN \ll 1$. Since typical fluid velocities $\mathbf{u}$ are $\mathcal{O}(1)$, we assume that the boundary is in a quasi-steady state, and $\mathbf{u}=0$ is a good approximation of the no-slip condition.


\subsection{Gibbs-Thomson effect}

During melting, solidification, and dissolution processes, surface effects such as surface tension and molecular kinetics modify the dynamics of the interface. The classical Gibbs-Thomson effect \cite{perez2005gibbs, moore2013self} is an idealized way to summarize these surface effects: an additional term that dissipates high curvature regions is added to the normal boundary velocity,
\begin{equation}
\label{gibbs}
V_n = \frac{\beta}{\PeN}\frac{\partial c}{\partial n} - \epsilon \left(\kappa^* - \frac{2\pi}{L}\right)  \quad \mbox{ on  } \Gamma,
\end{equation}
where $\kappa^*$ is the planar curvature and $L$ is the total arclength of $\Gamma$. The Gibbs-Thomson effect causes faster dissolution to occur in regions where the local curvature $\kappa^*$ is higher than a mean curvature $2\pi/L$ \cite{perez2005gibbs,moore2013self}. With no phase-change driven boundary velocity [that is, the first term on the right-hand side in \cref{gibbs}], the Gibbs-Thomson effect simply drives the boundary to a steady form: a circular arc with curvature $\kappa^* = 2\pi / L$ everywhere. Physically, the Gibbs-Thomson effect comes from the balance between the energy of fusion -- which represents the energy exchange during phase change -- and the surface energy that is a function of the curvature $\kappa^*$ \cite{perez2005gibbs}. When the local curvature is high enough, the surface energy overcomes the energy of fusion and drives interfacial motion. Numerically, the Gibbs-Thomson effect defines a minimal spatial scale at which Mullins-Sekerka instabilities occur.

\section{Numerical methods}
\label{methods}

Solving \cref{stefan-nc} presents two central difficulties: (1) solving the Navier-Stokes (NS) equations in the evolving domain $\Omega_\textnormal{liquid}$, and (2) solution of the advection-diffusion equation for $c$ in that same domain. Because the boundary evolution is itself driven by \emph{gradients} of $c$, the quantity $\partial c/\partial n$ must be captured accurately at the boundary. We first discuss the Navier-Stokes equations. In 2D, the NS equations can be reformulated using the stream function $\psi$ and the vorticity $\omega$, as
\begin{subequations}
\label{steamVor}
\begin{align}
     \frac{\partial \omega}{\partial t} + \mathbf{u}\cdot\nabla \omega =\frac{1}{\ReN} \Delta \omega - \frac{\partial c}{\partial x} &\quad \mbox{ in  } \Omega_{liquid},\label{streamEQN}\\
     \Delta \psi = -\omega,\ \mathbf{u} = \nabla_\bot \psi &\quad \mbox{ in  } \Omega_{liquid},\\
     \psi = \psi_n = 0 &\quad \mbox{ at  } \Gamma\cup\Gamma_w,
\end{align} 
\end{subequations}
where $\nabla_\bot = (\partial_y, -\partial_x)$. \tR{The driving term $-\partial c/\partial x$ arises due to the Boussinesq approximation of the buoyancy forces, which appears as the term $-c\hat{\mathbf{y}}$ in \cref{NSE-approx}.} In this formulation, solution of the pressure field is no longer required, which typically poses difficulties due to the lack of boundary conditions. Nevertheless, the equations do not decouple: \Cref{streamEQN} lacks a boundary condition on $\omega$, while there are two boundary conditions imposed on the stream function.

To efficiently discretize the full nonlinear system, we will treat the non-linear terms explicitly, and the linear terms implicitly. A full description of the timestepping scheme we use is given in \cref{numImp}; here we give a simplified description in order to introduce the various subproblems that must be solved. Representing the boundary $\Gamma$ with its Cartesian coordinates $\mathbf{X} = (x, y) \in \Gamma$, the simplest implicit-explicit time discretization is to combine the forward and backward Euler schemes:
\begin{subequations}
\begin{align}
 	 \left(\mathbb{I} - \frac{\Delta t}{\PeN} \Delta\right) c^{t+\Delta t} = c^{t} - \Delta t (\mathbf{u}^{t}\cdot\nabla c^{t}) &\quad \mbox{ in  } \Omega_{liquid},\\
     \left(\mathbb{I} - \frac{\Delta t}{\ReN} \Delta\right) \omega^{t+\Delta t} = \omega^{t} - \Delta t \left[\mathbf{u}^{t}\cdot\nabla \omega^{t} + \left(\frac{\partial c}{\partial x}\right)^{t+\Delta t}\right] &\quad \mbox{ in  } \Omega_{liquid},\\
     \Delta \psi^{t+\Delta t} = - \omega^{t+\Delta t},  \mathbf{u}^{t+\Delta t} = \nabla_\bot \psi^{t+\Delta t} &\quad \mbox{ in  } \Omega_{liquid},\\
     \mathbf{X}^{t+\Delta t} = \mathbf{X}^{t} + \Delta t V_n \mathbf{n} &\quad \mbox{ on  } \Gamma.
\end{align} 
\end{subequations}
with boundary conditions on $\psi^{t+\Delta t}$ and $c^{t+\Delta t}$ as given in \cref{stefan-nc,steamVor}. One can first evolve $c$, so that $c^{t+\Delta t}$ is known and can be used explicitly in the RHS of the vorticity equation. For more general, timestepping schemes, the temporal discretization takes the form:
\begin{subequations}
\label{NS-helmtz}
 \begin{align}
 	 \label{c-eqn}
 	 (\mathbb{I} - \sigma_c \Delta) c^{t+\Delta t} = f_c(c^t, \mathbf{u}^t) &\quad \mbox{ in  } \Omega_{liquid},\\
 	 \label{omega-eqn}
     (\mathbb{I} - \sigma_\omega \Delta) \omega^{t+\Delta t} = f_\omega(\mathbf{u}^t, \omega^t, c^{t+\Delta t}) &\quad \mbox{ in  } \Omega_{liquid},\\
     \label{psi-eqn}
     \Delta \psi^{t+\Delta t} = - \omega^{t+\Delta t} &\quad \mbox{ in  } \Omega_{liquid},\\
     \label{X-eqn}
     \mathbf{X}^{t+\Delta t} = f_X (\mathbf{X}^{t}, \mathbf{X}^{t+\Delta t}, V_n)  &\quad \mbox{ on  } \Gamma.
\end{align} 
\end{subequations}
Here $\sigma_c,\ \sigma_\omega,\ f_c$, and $f_\omega$ depend on the choice of time discretization. There are thus two essential components required for evolving this system: (1) a scheme for evolving the interface $\Gamma$ [\cref{X-eqn}], which will be introduced in \cref{section:theta_L}, and (2) Modified-Helmholtz solvers for both the concentration evolution \cref{c-eqn} and the coupled Navier-Stokes system [\cref{omega-eqn,psi-eqn}]. The methodology for solving these PDE will be introduced in \cref{section:ibse:helmholtz,section:ibse:stokes_coupled}, as well as some technical considerations that arise when solving the fully coupled system.

\subsection{$\theta$-$L$ method for boundary evolution}
\label{section:theta_L}

For a variety of boundary evolution problems, including the Stefan problem studied here, as well as problems with an interfacial surface tension, discretization and evolution of the boundary in Cartesian coordinates can lead to stability issues due to the fact that only the normal motion of the boundary $\Gamma$ is fixed by the physics. Discretized boundary markers moved only according to the normal velocity typically clump or spread. To alleviate the issues that arise with a direct discretization of the boundary position, we instead use the $\theta$-$L$ method to discretize the boundary by the arclength $s$ and tangent angle $\theta(s,t)$. This method is well-developed and has been used extensively for the study of 2D free-boundary problems in fluid dynamics \cite{hou1994removing, alben2002drag, moore2013self, shelley2011flapping,  ganedi2018equilibrium}. We provide a brief overview of the method here.

Consider a domain as depicted in \cref{fig1}(a) where the boundary $\Gamma$ has total arclength $L$. Given an arclength parameter $s$, we define a rescaled arclength $\alpha=s/L$, with $\alpha\in[0,1]$, where $\alpha=0$ corresponds to the top point. The boundary $\Gamma$ can be parameterized by the tangent angle $\theta(\alpha,t)$, the total arclength $L$, and the position of the top point, with the Cartesian coordinate recovered by solving $\partial_\alpha \mathbf{X} = (\partial_\alpha x, \partial_\alpha y) = (L \cos{\theta}, L \sin{\theta})$ and $\mathbf{X}(0) = (x_0, y_0)$ as the top point of the solid body. Evolution of the boundary $\mathbf{X} = (x(\alpha), y(\alpha))$ is given by
\begin{equation}
\label{X}
\frac{\partial \mathbf{X}}{\partial t} = V_n \mathbf{n} + V_s \mathbf{s},
\end{equation}
where $\mathbf{n}$ and  $\mathbf{s}$ are unit vectors along the normal and tangential direction at a given point on $\Gamma$. As the normal velocity $V_n$ governs the shape evolution, the actual shape dynamics of $\Gamma$ are independent of the tangential velocity $V_s$. We are thus free to add an arbitrary tangential velocity without altering the effective physics. One choice that preserves the equal-arclength distribution and keeps $\alpha$ and $t$ as independent variables \cite{moore2013self} is
\begin{equation}
\label{vs}
V_s = \int_0^\alpha \frac{\partial \theta}{ \partial \alpha'} V_n(\alpha')d\alpha' - \alpha \int_0^1 \frac{\partial \theta}{ \partial \alpha} V_n(\alpha')d\alpha'.
\end{equation}

With $V_n$ and $V_s$ specified, \cref{X} can be written as
\begin{subequations}
\label{theta-L}
\begin{align}
\frac{dL}{dt} &= -\int_0^1 \frac{\partial \theta}{ \partial \alpha} V_n(\alpha)d\alpha, \\
\frac{\partial \theta}{\partial t} &= \frac{1}{L}\left(\frac{\partial V_n}{\partial \alpha} + V_s \frac{\partial \theta}{\partial \alpha}\right). 
\end{align}
\end{subequations}

In addition to evolving $\theta(\alpha,t)$ and $L(t)$, the motion of an anchor point is required to recover the absolute location of the boundary $\Gamma$. In this paper, we use the top point $(x_0, y_0)$ as this moving anchor. These satisfy the equation
\begin{subequations}
\label{xy0}
	\begin{align}
		\frac{d x_0}{d t}	&= V_n(0)\cos{\theta(t,0)},	\\
		\frac{d y_0}{d t}	&= V_n(0)\sin{\theta(t,0)}. 
	\end{align}
\end{subequations}

When the interface $\Gamma$ is a smooth, closed curve, \cref{theta-L} can be solved accurately using a Fourier spectral method. When including the Gibbs-Thomson effect [\cref{gibbs}] in the boundary evolution, the $\theta$ equation is instead
\begin{equation}
\label{theta}
\frac{\partial \theta}{\partial t} = \frac{1}{L}\left(\frac{\beta}{\PeN}\frac{\partial^2 c}{\partial n \partial \alpha} + V_s \frac{\partial \theta}{\partial \alpha}\right) + \frac{\epsilon}{L^2}\frac{\partial^2 \theta}{\partial \alpha^2}.
\end{equation}

Thus the Gibbs-Thomson effect smooths the boundary by dissipating regions with high curvature.

\subsection{IBSE $^k$ for solving Poisson and Modified-Helmholtz equations}
\label{section:ibse:helmholtz}

In this section we provide a basic outline of the Immersed Boundary Smooth Extension (IBSE) method, focusing on the details most relevant to the numerical scheme for the Stefan and dissolution problems studied in this paper. A careful analysis of the method is presented in \cite{Stein2016252, Stein2017155}. \tR{The essential idea of this method is to find a smooth extension of the unknown solution from the physical domain to the entirety of a computationally simple domain, such that the first several derivatives of the unknown solution and its extension match across the physical boundary.} To simplify the presentation, we first consider the inhomogeneous Poisson problem:
\begin{subequations}
	\label{stokeseq:poisson}
	\begin{align}
		\Delta u  &=	f	&&\text{in }\Omega,	\\
		u		  &=	g	&&\text{on }\Gamma.
	\end{align}
\end{subequations}

One way to compute an inhomogeneous solution is to compute an extension of $f$ from $\Omega$ to a larger domain $C$ containing $\Omega$, which we assume has a simple geometry. Perhaps the simplest choice of $C$, and the one used throughout this paper, is the periodic rectangle $C=\mathbb{T}^2=[0, 2\pi]\times[0,2\pi]$, or a rescaled version of that domain. For this choice of $C$ the Poisson operator may be rapidly inverted via the fast-Fourier transform. The physical domain of the equations is $\Omega$, the exterior of that domain is $E$, and $C = \Omega \cup E$. Given a general extension $f^e\in C^{k}(C)$, $u$ satisfies \cref{stokeseq:poisson}, up to a homogeneous correction, to $\mathcal{O}(h^{k+1})$ where $h = 2\pi/N$ is the spatial resolution. There is significant freedom in choosing $f^e$. One such choice is to define $f^e=\Delta u^e$, where $u^e$ is an extension to the solution $u$ of the PDE. In this case, $u$ satisfies the PDE \emph{without} any homogeneous correction, although the extension $f^e$ must be determined implicitly, as the solution $u$, and hence its extension $u^e$, is not known \emph{a priori}. We describe some of the essential elements for computing $u$ in this manner below.

Let $\Omega\in C$ be compact and simply connected, with smooth boundary $\Gamma=\partial\Omega$. To facilitate generating the smooth extension $u^e$ of the solution $u$, as well as to impose boundary conditions on $\Gamma$, we will need to be able to communicate singular and hyper-singular force distributions known on $\Gamma$ to the grid; and to interpolate values and derivatives of functions known on the grid to the discrete boundary nodes of $\Gamma$. Communication of singular forces is done through the \emph{spread} operators $S_{(j)}$ as 
\begin{align}
	\label{stokeseq:derivative_spread_operator}
	(S_{(j)}F)(x) = (-1)^j\int_\Gamma F_j(\alpha)\frac{\partial^j\delta(x-X(\alpha))}{\partial n^j}\,dX(\alpha),
\end{align}
while approximations of a function and its derivatives at $\Gamma$ are computed via the \emph{interpolation} operators $S^*_{(j)}$:
\begin{align}
	\label{stokeseq:derivative_interpolation_operator}
	(S_{(j)}^*\xi)(\alpha) = (-1)^j\int_C \xi(x)\frac{\partial^j\delta(x-X(\alpha))}{\partial n^j}\,dx
\end{align}
As is suggested by the notation, the spread and interpolation operators for the $j^\text{th}$ derivatives are adjoint to each other, and for $j=0$ these operators reduce to the classical Immersed Boundary (IB) spread and interpolation operators \cite{peskin2002immersed}. \tR{The interpolation operator $S_{(j)}^*$ maps the $j^\text{th}$ normal derivative of a function $\xi$ from the domain $C$ to the boundary $\Gamma$, while the spread operator $S_{(j)}$ maps the hyper-singular forces from the boundary $\Gamma$ back to the domain $C$.} To simplify the notation for the rest of this study, we define the composite operators $T_k,\ T^*_k,$ and $R^*_k$ as
\begin{subequations}
	\label{eq:define_tks}
	\begin{align}
		T_k &= \sum_{j=0}^{k}S_{(j)},	\\
		T^*_k &= 
		\begin{pmatrix}
			S_{(0)}^*	&	S_{(1)}^*	&	\cdots	&	S_{(k)}^*
		\end{pmatrix}^\intercal,\\
		R^*_k &= 
		\begin{pmatrix}
			S_{(1)}^*	&	\cdots	&	S_{(k)}^*
		\end{pmatrix}^\intercal.
	\end{align}
\end{subequations}

The operator $T_k^*$ interpolates a function and its first $k$ normal derivatives to the boundary; $R_k^*$ provides an interpolation of the first $k$ normal derivatives to the boundary, excluding the function value itself; the operator $T_k$ spreads a set of singular forces ($\delta$-like) and hyper-singular forces  (like the first $k$ normal derivatives of the $\delta$-function) from the boundary to the domain.


To solve \cref{stokeseq:poisson}, a function $\eta$ is found that smoothly extends the unknown solution $u$ from $\Omega$ to $C$ such that $\eta\in C^k(C)$ and the first $k$ derivatives of $u$ match the first $k$ derivatives of $\eta$ at the boundary: $\partial_j u /\partial^j n= \partial_j \eta/\partial^j n$ for $j = 1,\ldots, k$ on $\Gamma$. There are many such extensions. We choose to compute $\eta$ by solving a high-order PDE of the form $\mathcal{H}^k\eta = 0$ in $C$, where $\mathcal{H}^k$ is a differential operator with sufficient order to allow us to impose that the derivatives of $\eta$ at the boundary match those of $u$. One such choice of $\mathcal{H}^k$ is the polyharmonic operator $\Delta^{k+1}$; this choice has a nullspace and is poorly conditioned; the specific operator $\mathcal{H}^k$ that we use will be defined in \cref{numSec}. The function $\eta$ then serves to define a force in all of $C$, given by $f^e = \chi_\Omega f + \chi_E\Delta\eta$. Inversion of the periodic Laplace operator with this force provides a solution $u_e$, smooth in $C$, that satisfies the boundary conditions and converges rapidly. The full formulation, which we will refer to as the IBSE-$k$ system, is given by:
\begin{subequations}
	\label{stokeseq:ibse_poisson}
	\begin{align}
		\Delta u_e - \chi_E\Delta\eta	&=	\chi_\Omega f	&&\text{in }C,	\label{stokeseq:ibse_poisson:a}	\\
		\mathcal{H}^k{\eta} + T_k F				&=	0							&&\text{in }E,	\label{stokeseq:ibse_poisson:b}	\\
		R_k^*\eta	&=	R_k^*u_e,	\label{stokeseq:ibse_poisson:c}	\\
		S^*_{(0)}u_e		&=	g.	\label{stokeseq:ibse_poisson:d}
	\end{align}
\end{subequations}
In \cite{Stein2016252}, the IBSE-$k$ formulation given in \cref{stokeseq:ibse_poisson} has been verified to produce $C^k(C)$ solutions that converge at a rate of $O(\Delta x^{k+1})$ for $k = 1,2,3$; and may be used for solving the Helmholtz and Modified-Helmholtz equations, by replacing $\Delta$ with $(k^2+\Delta)$ and $(k^2-\Delta)$, respectively. In \cite{Stein2017155}, the IBSE method was extended to solve the Stokes and Navier-Stokes equations. In the following sections, we will discuss how to modify these methods to provide a solver for the Stefan and dissolution problems.

\subsection{Navier-Stokes equations coupled with concentration field}
\label{section:ibse:stokes_coupled}

In order to solve the coupled dissolution problem with the IBSE method, we let the physical domain be $\Omega = \Omega_{liquid}$ and the extension domain be $E=\Omega_{solid}$, and define the modified Helmholtz operators:
\begin{subequations}
	\label{hemholtz_ops}
	\begin{align}
		\mathcal{L}_\omega	&= (\mathbb{I} - \sigma_\omega \Delta),	\\
		\mathcal{L}_c	&= (\mathbb{I} - \sigma_c \Delta).
	\end{align}
\end{subequations}
Roughly speaking, these operators will be used to define solvers similar to the solver described in  \cref{section:ibse:helmholtz}. Specifically, for the concentration \cref{c-eqn}, we solve
\begin{subequations}
	\label{c-eqn:ibse_poisson}
	\begin{align}
		\mathcal{L}_c c - \chi_E\mathcal{L}_c\eta_c	&=	\chi_\Omega f_c	&&\text{in }C,	\label{c-eqn:ibse_poisson:a}	\\
		\mathcal{H}^k{\eta_c} + T_k F_c				&=	0							&&\text{in }E,	\label{c-eqn:ibse_poisson:b}	\\
		R_k^*(\eta_c-c)	&=	0 &&\text{at }\Gamma\cup \Gamma_w,	\label{c-eqn:ibse_poisson:c}	\\
		S^*_{(0)}c		&=	g &&\text{at }\Gamma, \label{c-eqn:ibse_poisson:d}\\
		S^*_{(1)}c		&=	0 &&\text{at }\Gamma_w.  \label{c-eqn:ibse_poisson:e}
	\end{align}
\end{subequations}

Next, we solve the Navier-Stokes equations, \cref{omega-eqn,psi-eqn}, with the known concentration field $c$, by solving \cref{c-eqn:ibse_poisson}. In \cite{Stein2017155}, the authors used a velocity-pressure formulation for solving fluid problems. Importantly, they noted that the best stability and error was achieved when the regularity of extensions matched those expected from derivative counting the PDEs: that is, if $\mathbf{u}$ is extended to be $C^k(C)$, then $p$ should only be extended to be in $C^{k-1}(C)$. We find a similar phenomena in the streamfunction-vorticity formulation, and hence to find a velocity $\mathbf{u} = (u,v)$ in $C^k(C)$, we choose to extend $\omega\in C^{k-1}(C)$ and $\psi\in C^{k+1}(C)$. Accordingly, the appropriate formulation is 
\begin{subequations}
	\label{omega-eqn:ibse_stokes}
	\begin{align}
		\mathcal{L}_\omega \omega - \chi_E\mathcal{L}_\omega\eta_\omega	&=	\chi_\Omega f_\omega	&&\text{in }C,	\label{omega-eqn:ibse_poisson:a}	\\
		\Delta \psi - \chi_E\Delta\eta_\psi	+\omega &= 0	&&\text{in }C,	\label{omega-eqn:ibse_poisson:b}	\\
		\mathcal{H}^{k-1}{\eta_\omega} + T_{k-1} F_\omega				&=	0							&&\text{in }E,	\label{omnega-eqn:ibse_poisson:c}	\\
		\mathcal{H}^{k+1}{\eta_\psi} + T_{k+1} F_\psi				&=	0							&&\text{in }E,	\label{omnega-eqn:ibse_poisson:d}	\\
		R_{k-1}^*(\eta_\omega-\omega)	&=	0 &&\text{at }\Gamma\cup \Gamma_w, 	\label{omega-eqn:ibse_poisson:e}	\\
		R_{k+1}^*(\eta_\psi-\psi)	&=	0 &&\text{at }\Gamma\cup \Gamma_w,	\label{omega-eqn:ibse_poisson:f}	\\
		S^*_{(0)}\psi		&=	0 &&\text{at }\Gamma\cup \Gamma_w,	\label{omega-eqn:ibse_poisson:g}\\
		S^*_{(1)}\psi		&=	0 &&\text{at }\Gamma\cup \Gamma_w.	\label{omega-eqn:ibse_poisson:h}
	\end{align}
\end{subequations}

The unknowns are $(\omega, \psi,\eta_\omega, \eta_\psi, F_\omega, F_\psi)$. In the next section, we discuss details of the numerics.

\section{Numerical implementation}
\label{numImp}
\subsection{Numerical scheme for the $\theta-L$ method}

On the boundary $\Gamma$, the rescaled arclength $\alpha\in[0,1)$ is discretized using equispaced points, and Fourier spectral (or pseudo-spectral) methods are used. The explicit equations in \cref{theta-L,xy0} are integrated with a fourth-order Adam-Bashforth method. At time $t = m\Delta t$, \cref{theta} can be split into the non-stiff explicit part $\mathcal{E}^m = (1/L^m) [(\beta/  \PeN)(\partial^2 c/\partial n \partial \alpha)^m + V_s^m (\partial \theta/\partial \alpha)^m ]$ and stiff implicit part $\mathcal{I}^m = [\epsilon/(L^m)^2](\partial^2 \theta/\partial \alpha^2 )^m.$ For this equation, we use the fourth-order implicit-explicit (IMEX) Backward Differentiation formula and the fourth-order Adam-Bashforth formula:
\begin{align}
\label{theta-IMEX}
\mathbf{L}^{m+1} &= \mathbf{L}^m + \frac{\Delta t}{24}\left(55 \mathbf{V}_L^m - 59 \mathbf{V}_L^{m-1} + 37 \mathbf{V}_L^{m-2}-9\mathbf{V}_L^{m-3}\right),\\
\frac{25}{12}\theta^{m+1} - 4\theta^m + 3\theta^{m-1} - \frac{4}{3}\theta^{m-2} + \frac{1}{4}\theta^{m-3} &= \Delta t \left[\mathcal{I}^{m+1} + 4\mathcal{E}^{m} - 6\mathcal{E}^{m-1} + 4\mathcal{E}^{m-2} - \mathcal{E}^{m-3}\right].
\end{align}

Here $\mathbf{L}^m = (L^m,x_0^m,y_0^m)$ and $\mathbf{V}_L^m = \left(-\int_0^1 \frac{\partial \theta}{\partial \alpha}(m\Delta t,\alpha) V_n(m\Delta t,\alpha)d\alpha, V_n(m\Delta t,0)\cos{\theta(m\Delta t,0)},V_n(m\Delta t,0)\sin{\theta(m\Delta t,0)}\right) $.  At timestep $m+1$, the evaluation of $\mathcal{I}^{m+1}$ would require $L^{m+1}$, so the explicit integration of $x_0^{m+1}, x_0^{m+1}$ and $L^{m+1}$ is performed first. $\theta^{m+1}$ is then found via the Fourier method.

\subsection{Discretization of spread, interpolation and extension operators}
\label{numSec}

In this section, we define the discrete spread, interpolation and extension operators introduced in \cref{section:ibse:helmholtz}; we will not explicitly distinguish the discretized operators through different notation. Let $\tilde{\delta}$ denote a regularized $\delta$-function, defined by Cartesian products of regularized one-dimensional $\delta$-functions. We define normal derivatives of $\tilde{\delta}$ to be
\begin{equation}
	\frac{\partial^j\tilde{\delta}}{\partial n^j} = n_{i_1}\cdots n_{i_{j}}\frac{\partial^j\tilde{\delta}}{\partial x_{i_1}\cdots \partial x_{i_j}},
	\label{stokeseq:normal_derivative_formula}
\end{equation}
where repeated indices $i_j$ are summed according to the Einstein summation convention. Using the standard spectral discretization of the integral in \cref{stokeseq:derivative_interpolation_operator}, and replacing the $\delta$ function with its regularized equivalent, we define the discretized spread operator to be:
\begin{equation}
\label{S-operator}
	(S_{(j)}F)(\mathbf{x}) = \sum_{i=1}^{n_\text{bdy}} F(\alpha_i)\frac{\partial^j\tilde\delta(\mathbf{x}-\mathbf{X}_i)}{\partial n^j}\Delta s,
\end{equation}
$\mathbf{X}_i = \mathbf{X}(\alpha_i)$ is the $i$th coordinate of the boundary and $\Delta s = L / n_{bdy}$ (note that for arbitrary parametrizations, $\Delta s$ depends on $\sqrt{X_s^2 + Y_s^2}$; however, since we use an arclength parametrization, this is constant). The number of nodes in the quadrature is chosen so that $\Delta s \approx 2 \Delta x$.  Choosing $\Delta s$ smaller yields better accuracy while choosing it larger yields better numerical stability. The number of boundary points is $n_{bdy} \approx L / (2 \Delta x)$. Note that during the process of melting or dissolving, the number of boundary points $n_{bdy}$ must thus be updated as the total arclength $L$ changes. 

The interpolation operator is then defined through the adjoint property $\left<u,S_{(j)}F\right>_C=\left<S_{(j)}^*u,F\right>_\Gamma$, and is explicitly given by
\begin{equation}
\label{ST-operator}
	(S_{(m)}^*u)(\alpha_k) = \sum_{i=1}^N \sum_{j=1}^N u_{ij}\frac{\partial^j\tilde\delta(x_{ij}-\mathbf{X}_k)}{\partial n^m}\,\Delta x \Delta y.
\end{equation}
Here the vector $x_{ij}-\mathbf{X}_k = (2\pi i/N - X_{k,1} , 2\pi j/N - X_{k,2})$ points from the boundary point $\mathbf{X}_k \in \Gamma$ to the domain point $x_{ij} \in C$; the gridspacing $\Delta x = \Delta y = 2\pi/N$.

The accuracy of these operators is analyzed in \cite{Stein2016252}, and depends on the choice of the underlying regularized $\delta$-function. We use the $C^3$ function with a support width of $16\Delta x$ defined in \cite{Stein2016252}; with this choice the interpolation operator provides fourth-order accurate approximations for the $0$th to $3$rd normal derivatives of smooth functions.

Finally, we must choose the extension operator(s) $\mathcal{H}^k$ used in \cref{c-eqn:ibse_poisson,omega-eqn:ibse_stokes}. This should be a high-order differential operator, so that a sufficient number of boundary conditions can be imposed. An obvious choice is the poly-harmonic operator of the appropriate order, but this is both poorly conditioned and has a null-space. Adding a scalar $\Theta$ remedies the nullspace, and its size may be used to control the condition number of the system that must be solved, at the expense of adding an artificial length-scale to the problem that must be resolved by the discretization:

\begin{equation}
	\label{stokeseq:Helmholtz_definition}
	\mathcal{H}^k = \Delta^{k+1} + (-1)^{k+1}\Theta(k,m).
\end{equation}

The choice of $\Theta$ depends on the smoothness $k$ and the largest wave-number $m$ present in the discrete Fourier transform on the discrete domain $C_E$. The condition number of $\mathcal{H}^k$ is
\begin{equation}
	\kappa = 1 + \frac{m^{2(k+1)}}{\Theta},
\end{equation}
while the intrinsic lengthscale introduced is $\Theta^{-1/2(k+1)}$.  We choose $\Theta$ as
\begin{equation}
	\label{stokeseq:Theta_definition}
	\Theta^* = \left(\frac{1}{\mathcal{N}\Delta x}\right)^{2(k+1)},
\end{equation}
as in \citep{Stein2017155}. Here $\mathcal{N}$ is a parameter that controls how many points are used to resolve the intrinsic lengthscale introduced by $\mathcal{H}^k$; for all simulations in this paper $\mathcal{N}=10$ is used.

\subsection{Smoothed characteristic functions $\hat{\chi}_E$ and $\hat{\chi}_\Omega$ for stiff modified-Helmholtz problems}
When solving the diffusion equation with a small diffusion coefficient and/or small timestep, time-discretization yields a modified Helmholtz problem with a very small Helmholtz parameter. In fact, this parameter is typically \emph{artificially small}, in that it introduces a length-scale that is below any relevant to the physical problem. However, if this length-scale is not resolved, the standard IBSE solver may yield low-quality solutions, especially during brief transient phases (i.e. for small times $t$ around the startup of a dissolution problem, when $c$ jumps discontinuously from $1$ to $0$ across the interface, and diffusion has yet to have enough time to regularize the solution). Stability, although not accuracy, may be recovered in these circumstances by using smoothed, rather than sharp, characteristic functions $\hat{\chi}_\Omega$ and $\hat{\chi}_E$. The family of functions we adopt here are the Wendland functions \cite{chernih2014wendland}:

\begin{equation}
\label{wendland}
\phi_{l,m}(r) = \left\{
\begin{array}{cl}
\frac{1}{\Gamma(m) 2^{m-1}} \int_r^1 s(1-s)^l(s^2-r^2)^{m-1}ds &\mbox{for } 0\leq r\leq 1, \\
0  &\mbox{for } r>1.
\end{array}
\right.
\end{equation}

Here $m$ is an integer controlling the smoothness of Wendland function, and $l = \floor*{m+d/2} + 1$ with $d=2$ as the spatial dimension in our study. It can be shown that \cref{wendland} produces a $C^{2m}(\mathbb{R}^+)$ function. We also define the integral of Wendland function as $\Phi(r) = \int_{-\infty}^r \phi_{l,m}(|s|)ds / \int_{-\infty}^\infty \phi_{l,m}(|s|)ds$, so $\Phi(r) = 0$ when $r < -1$ and $\Phi(r) = 1$ when $r>1$.

Throughout our study, we choose $m = 2$ and $l = 4$ so all functions involved are at least $C^{4}$, and define the smoothed characteristic functions $\hat{\chi}_E$ and $\hat{\chi}_\Omega$ as 
\begin{align}
\label{chihat}
\hat{\chi}_\Omega (\mathbf{x}) &=  \Phi( r^+(\mathbf{x}) / d_s), \\
\hat{\chi}_E (\mathbf{x}) &= 1 - \hat{\chi}_\Omega (\mathbf{x}).
\end{align}
Here $r^+(\mathbf{x})$ is the signed distance between $\mathbf{x}$ and the boundary $\Gamma$, such that $r^+(\mathbf{x}) > 0$ when $\mathbf{x} \in \Omega$ and $r^+(\mathbf{x}) < 0$ when $\mathbf{x} \in E$. $d_s$ is a smoothing length-scale that is chosen to be $4 \Delta x$. In general, $d_s$ should be small enough that the boundary layer structure can be sufficiently resolved.

\subsection{Inversion of the IBSE-k system using Schur complements}
\label{Schur-section}
For the pure Stefan problem, \cref{c-eqn:ibse_poisson} can be written in block form as
\begin{equation}
	\label{Helmholtz:the_discrete_system_matrix}
	\left(
	\begin{array}{cc|c}
		\mathcal{L}_c			&	-\hat{\chi}_E\mathcal{L}_c & \mathbf{0}\\
		\mathbf{0} &	\mathcal{H}^k		&	T_k\\
		\hline
			R_k^*	&	-R_k^* & \mathbf{0}	\\
			S^*_{(0)}	& \mathbf{0}	 & \mathbf{0}\\
			
	\end{array}
	\right)
	\begin{pmatrix}
		c	\\	\eta_c	\\ \hline	\multirow{2}{*}{$F_c$} \\ \\
	\end{pmatrix}
	=
	\begin{pmatrix}
		\hat{\chi}_\Omega f_c	\\
			\mathbf{0}   \\	\hline
			\mathbf{0}	\\
			g	\\
	\end{pmatrix}.
\end{equation}
As shown in \citep{Stein2016252, Stein2017155}, the size of this system can be reduced to $k n_\text{bdy}\times k n_\text{bdy}$ by computing the Schur complement ($\SC$)
\begin{equation}
	\SC =
	\begin{pmatrix}
		R_k^*	&	-R_k^*	\\
		S^*_{(0)} & \mathbf{0}
	\end{pmatrix}
	\begin{pmatrix}
		\mathcal{L}_c	&	-\hat{\chi}_E\mathcal{L}_c	\\
		\mathbf{0}			&	\mathcal{H}^k
	\end{pmatrix}^{-1}
	\begin{pmatrix} \mathbf{0} \\
	T_k
	\end{pmatrix}.
	\label{eq:the_schur_complement}
\end{equation}
\Cref{Helmholtz:the_discrete_system_matrix} may then be reduced to
\begin{equation}
	\SC \cdot F_c
	=
	\begin{pmatrix}
			R^*_{k}	\mathcal{L}_c^{-1}
	\hat{\chi}_\Omega f_c\\ S^*_{(0)} \mathcal{L}_c^{-1}
	\hat{\chi}_\Omega f_c -g	
	\end{pmatrix}.
	\label{Helmholtz:schur_system}
\end{equation}

During each time step, the $m$-th column of $\SC$ can be prepared by applying the RHS of \cref{eq:the_schur_complement} to a vector $\mathbf{a}$ such that $a_j = \delta_{jm}$. Although this computation is relatively expensive and must be done $k n_\text{bdy}$ times, it is embarrassingly parallel, significantly reducing the preparation cost of SC on a multicore system. \Cref{Helmholtz:schur_system} can then be inverted to solve for $F_c$. Once $F_c$ is known, we may find $\eta_c$ and $c$:
\begin{subequations}
\label{c_eta_eqn}
\begin{align}
\eta_c &= (\mathcal{H}^k)^{-1} T_k F_c, \\
c &= \mathcal{L}_c^{-1} (\hat{\chi}_\Omega f_c + \hat{\chi}_E \mathcal{L}_c \eta_c ).
\end{align}
\end{subequations}

It is worth noting that $\SC$ depends only on the boundary geometry, which is slowly changing in the Stefan problem. As the preparation of the $\SC$ is expensive, we seek to reuse it, effectively amortizing its formation cost over a number of timesteps. In the simulation, $\SC^{(m)}$ is prepared at a time step $m$ and decomposed into LU components such that $L^{(m)}U^{(m)} = \SC^{(m)} $. Since the boundary $\Gamma$ changes slowly, $M = (L^{(m)}U^{(m)})^{-1} $ serves as a good preconditioner for solving the linear system \cref{Helmholtz:schur_system} for some number of timesteps. Once the $\SC$ fails to serve as an effective preconditioner, it is recomputed. Reformation of the $\SC$ is triggered by two criteria: (i) it takes too many iterations for GMRES to converge (more than 10 iterations in all examples in this manuscript); (ii) the boundary size $n_{bdy}$ is changed (since $n_{bdy} \approx L / 2 \Delta x$ and the arclength $L$ is changing). We call recomputation of Schur complement a ``$\SC$ renewal''. In \cref{highRa}, we show how often $\SC$ renewal happens for a particular simulation.

Next, the Schur complement for \cref{omega-eqn:ibse_stokes} is formed in the similar way. The discrete equation is
\begin{equation}
	\label{stokeseq:the_discrete_system_matrix}
	\left(
	\begin{array}{cccc|cc}
		\mathcal{L}_\omega	&	\mathbf{0}	&	-\hat{\chi}_E\mathcal{L}_\omega	&	\mathbf{0}			&	\mathbf{0}	&	\mathbf{0}	\\
		\mathbb{I}			&	\Delta		&	\mathbf{0}						&	-\hat{\chi}_E\Delta	&	\mathbf{0}	&	\mathbf{0}	\\
		\mathbf{0}			&	\mathbf{0}	&	\mathcal{H}^{k-1}				&	\mathbf{0}			&	T_{k-1}		&	\mathbf{0}	\\
		\mathbf{0}			&	\mathbf{0}	&	\mathbf{0}						&	\mathcal{H}^{k+1}	&	\mathbf{0}	&	T_{k+1}		\\
		\hline
		R_{k-1}^*	&	\mathbf{0}	&	-R_{k-1}^*	&	\mathbf{0}	&	\mathbf{0}	&	\mathbf{0}	\\
		\mathbf{0}	&	R_{k+1}^*	&	\mathbf{0}	&	-R_{k+1}^*	&	\mathbf{0}	&	\mathbf{0}	\\
		\mathbf{0}	&	T^*_1		&	\mathbf{0}	&	\mathbf{0}	&	\mathbf{0}	&	\mathbf{0}	\\
	\end{array}
	\right)
	\begin{pmatrix}
		\omega	\\	\psi	\\	\eta_\omega	\\	\eta_\psi	\\ \hline	\multirow{2}{*}{$F_\omega$}	\\	\multirow{2}{*}{$F_\psi$} \\ \\
	\end{pmatrix}
	=
	\begin{pmatrix}
		\hat{\chi}_\Omega f_\omega	\\ \mathbf{0}\\\mathbf{0}	\\	\mathbf{0}	\\	\hline
			\mathbf{0}	\\
			\mathbf{0}	\\
		\mathbf{0}
	\end{pmatrix}.
\end{equation}

To distinguish from $\SC$, we define the Grand Schur complement ($\GSC$) for this system,

\begin{equation}
	\label{stokeseq:define_schur_complement}
	\GSC = 
	\begin{pmatrix}
		R_{k-1}^*	&	\mathbf{0}	&	-R_{k-1}^*	&	\mathbf{0}	\\
		\mathbf{0}	&	R_{k+1}^*	&	\mathbf{0}	&	-R_{k+1}^*	\\
		\mathbf{0}	&	T^*_1		&	\mathbf{0}	&	\mathbf{0}
	\end{pmatrix}
	\begin{pmatrix}
		\mathcal{L}_\omega	&	\mathbf{0}	&	-\hat{\chi}_E\mathcal{L}_\omega & 	\mathbf{0}			\\
		\mathbb{I}			&	\Delta		&	\mathbf{0}						&	-\hat{\chi}_E\Delta	\\
		\mathbf{0}			&	\mathbf{0}	&	\mathcal{H}^{k-1}				&	\mathbf{0}			\\
		\mathbf{0}			&	\mathbf{0}	&	\mathbf{0}						&	\mathcal{H}^{k+1}			
	\end{pmatrix}^{-1}
	\begin{pmatrix}
		\mathbf{0}	&	\mathbf{0}	\\
		\mathbf{0}	&	\mathbf{0}	\\
		T_{k-1}		&	\mathbf{0}	\\
		\mathbf{0}	&	T_{k+1}	
	\end{pmatrix}.
\end{equation}

\Cref{stokeseq:the_discrete_system_matrix} becomes
\begin{equation}
	\bigg(		\quad \GSC \quad \bigg)
	\begin{pmatrix}
			F_\omega	\\	F_\psi
	\end{pmatrix}
	=
	\begin{pmatrix}
			R_{k-1}^*	&	\mathbf{0}	\\
			\mathbf{0}	&	R_{k+1}^* 	\\
			\mathbf{0}	&	T^*_1
	\end{pmatrix}
	\begin{pmatrix}
		\mathcal{L}_\omega	&	\mathbf{0}	\\
		\mathbb{I} 			&	\Delta
	\end{pmatrix}^{-1}
	\begin{pmatrix}
		\hat{\chi}_\Omega f_\omega	\\ \mathbf{0}
	\end{pmatrix}.
	\label{stokeseq:schur_system}
\end{equation}

Once \cref{stokeseq:schur_system} has been solved for $(F_\omega, F_\psi)$, the remaining unknowns $(\omega, \psi, \eta_\omega, \eta_\psi)$ can be determined in a manner analogous to \cref{c_eta_eqn}. The GSC is a $2kn_{bdy} + 1$ square system, where the additional equation is added due to the null space of the Laplace operator (see Appendix B of \cite{Stein2016252}).

\subsection{Outline of the numerical solver}

Here we sketch a brief outline to solve the Stefan problem and its related flow problem with IBSE. At each time step,

\begin{enumerate}
\item The shape of the moving boundary is determined by solving \cref{xy0,theta-IMEX}. \fourth order Adam-Bashforth and IMEX schemes are used for time integration, and the spatial equation is solved using a Fourier spectral method. 

\item Operators such as $T, T^*, R, R^*, S, S^*$ are prepared according to \cref{eq:define_tks,S-operator,ST-operator}.

\item If the simulation is running for the first time, or one of the two $\SC$ ($\GSC$) refreshing conditions is met, the Schur complement will be prepared according to the procedure introduced in \cref{Schur-section}.

\item The concentration field $c$ is solved first through \cref{Helmholtz:the_discrete_system_matrix} - \cref{c_eta_eqn}. The value of $\partial_x c$ is then evaluated with Fourier differentiation, the value of $\partial_n c$ at the boundary is calculated by applying $S^*_{(1)}$ to $c$ and the boundary velocity $V_n$ is computed according to \cref{gibbs}. 

\item The vorticity $\omega$ and the stream function $\psi$ equations are solved through \cref{stokeseq:the_discrete_system_matrix} - \cref{stokeseq:schur_system}. The flow velocity field $\mathbf{u} = \grad_\perp \psi$ is then calculated and serves as the advection term $\mathbf{u}\cdot \grad c$ in $f_c$ for the next time step. 
\end{enumerate}

For a pure Stefan problem without flow, only step 1-4 are involved. 
\section{Results: Stefan problems}
\label{stefanSection}

In this section, we use the method described in \cref{methods,numImp} to solve Stefan problems with the form:
\begin{subequations}
\label{stefan-1}
\begin{align}
     \frac{\partial c}{\partial t} = \Delta c &\quad \mbox{ in  } \Omega_{liquid},\\
     V_n =  \beta \frac{\partial c}{\partial n} &\quad \mbox{ on  } \Gamma, \\
     c = 0 &\quad \mbox{ on  } \Gamma,\\
     c(\mathbf{x}, 0) = -1  &\quad \mbox{ in  } \Omega_{liquid}.
\end{align} 
\end{subequations}
Physically, these equations describe the growth of a solid body into its over-cooled liquid surrounding as discussed in \cref{intro2Stefan}. We first examine the convergence of the numerical scheme with an analytic solution, demonstrating the expected order of convergence for the temperature and boundary position (up to third-order, in $L^\infty$, depending on the regularity of extensions). We then examine an unstable problem, and show that the method qualitatively produces the typical growth patterns driven by the Mullins-Sekerka (MS) instability.

\subsection{Convergence of the Stefan solver}
\label{stefan:convergence}

\begin{figure}
	\centering
	\includegraphics[width=0.78\textwidth]{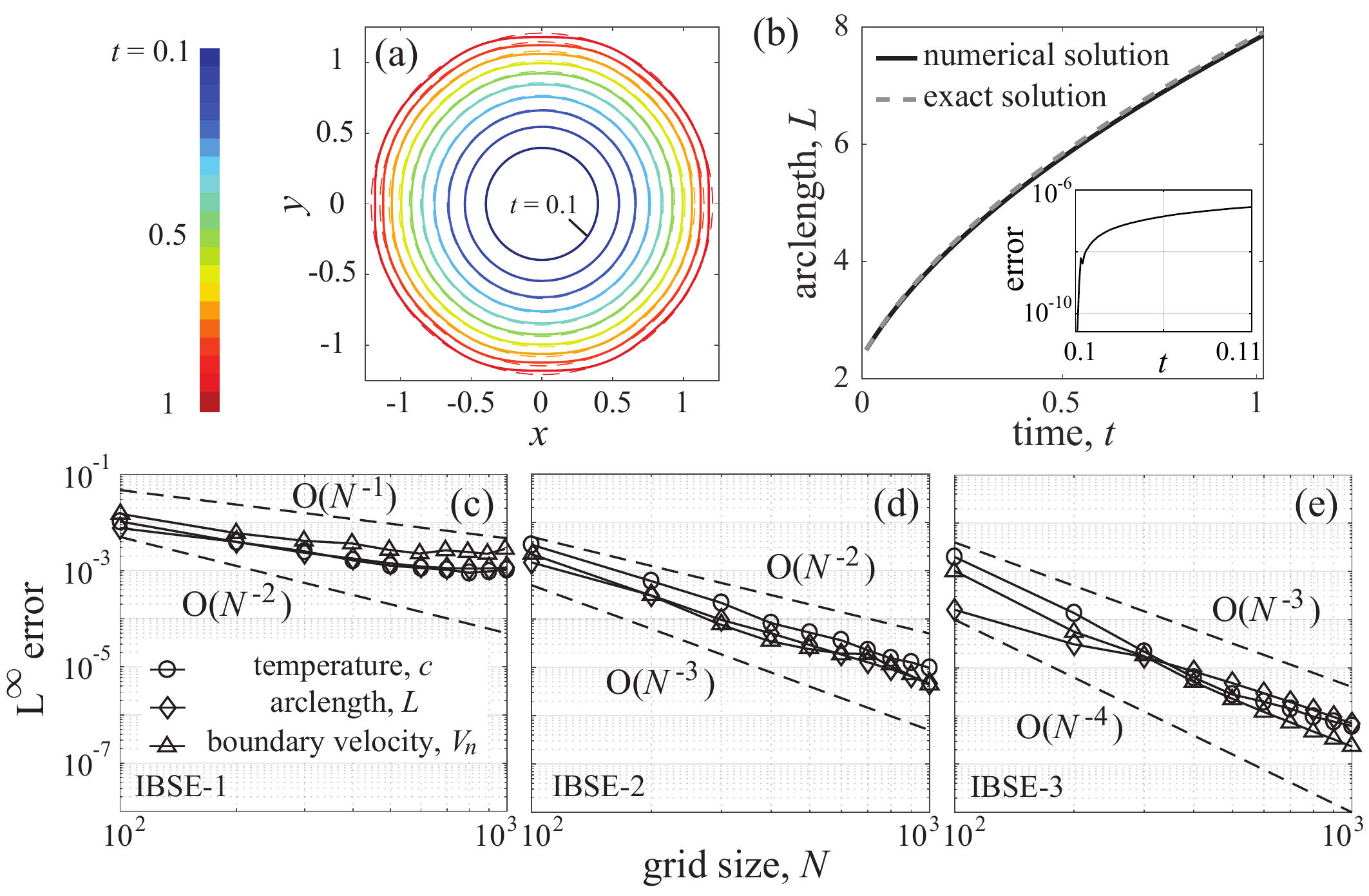}
	\caption{Frank disk solution of the Stefan problem and a convergence test of the numerical solver. (a) \tO{The liquid-solid interface given by the} numerical solution \tO{to the periodic problem} (\tO{IBSE-2, solid line}) and the exact solution \tO{to the free-space problem} (\tO{dashed line}). The circular initial shape remains circular until periodicity limits the growth of the interface. (b) The total arclength of the interface $L$ grows as the square root of time. The numerical solution (IBSE-2, solid line) matches well with the free-space analytic solution (dashed line), especially at short times. \tO{\textit{Inset:} Relative error of $L$ at early times. Data from IBSE-3 simulation with $N = 1000$.} (c)-(e) $L^\infty$ error of the numerical solutions. For IBSE-1, 2 and 3, the concentration field $c$, total arclength $L$ and boundary velocity $V_n$ converge to the exact solution at \first, \second, and \third order, respectively.}
	\label{fig2}
\end{figure}

An exact solution for a circle growing in infinite space (known as the ``Frank disk") \cite{Gibou2005577} stays circular during the freezing process (although small perturbations, if present, will drive the Mullins-Sekerka instability, as in \cref{stefan:instability}). The radius of the Frank disk, as a function of time, is $R(t) = S_0 \sqrt{t}$, where $S_0$ is a parameter determined by $\beta$ and $c_\infty$.  The associated temperature field is:
\begin{equation}
\label{exact_c}
c = \left\{
\begin{array}{cl}
0 &\mbox{for } s \leq S_0, \\
c_\infty \left(1-\frac{F(s)}{F(R_0)}\right)  &\mbox{for } s>S_0,
\end{array}
\right.
\end{equation}
where $s = |\mathbf{x}| / \sqrt{t}$, $F(s) = E_1(s^2/4)$, and $E_1(z) = 2 \int_z^\infty e^{-t}/t\, dt$. \tO{To analyze the validity and accuracy of our solver}, we let $\beta=-0.4$, $S_0=1.2$, and begin simulations starting at $t=0.1$ with an initial radius $R(0.1)=0.38$. \tO{In these simulations, as in all simulations throughout this manuscript, we use a Fourier pseudo-spectral method as the basis for the IBSE solver. Our system thus has naturally periodic boundary conditions, while the analytical form for the Frank disk solution is valid only in infinite space. The numerical solution to the periodic problem, as computed by the IBSE-$2$ method, together with the analytical Frank disk solution, is shown in \cref{fig2}(a). Despite the differing boundary conditions, the periodic solution and the analytic solutions agree well over short to medium times. As the interface diameter approaches the width of the periodic interval, periodicity pushes the interface shape away from the analytic circular solution. The arc-length of the liquid-solid interface is shown in \cref{fig2}(b), with the absolute difference between analytic and computed solutions shown in the inset for short times, where the interface is far from the periodic boundary. At these small times, this concordance is on a similar scale to the error we expect from our solver.} In order to test the accuracy of our solver, we thus solve for only a short amount of time: to $t=0.11$ (for times larger than $t\gtrsim 0.11$, periodic effects, although still small, become comparable with the errors achieved by the IBSE-$3$ method.). In \cref{fig2}(c)-(e), we show a convergence study for the arclength $L$ of the boundary, the concentration field $c(\mathbf{x}, t)$ and the boundary velocity $V_n$, as computed using the IBSE-$1$, $2$, and $3$ methods with $\Delta t=0.1/N$ and $N/10$ timesteps. Startup values for the BDF-based IMEX timestepping scheme are computed from the analytic solution. All errors are measured against the analytic solution in $L^\infty$, and a convergence rate of $\mathcal{O}(h^{k})$ is observed for all variables examined. \tO{Although these errors are a mixture of discretization and periodicity-related errors, the periodicity-related error must be dominated by our discretization error over these times, or we would have seen stagnating errors, rather than a $k$-th order convergence rate}. The IBSE-$k$ method achieves an $\mathcal{O}(h^{k+1})$ convergence rate for Dirichlet problems with defined boundaries \cite{Stein2016252}. In this problem the interface location is moving with the normal velocity $V_n$. Computing this velocity requires estimating $\partial c/\partial n$, which is only accurate to $\mathcal{O}(h^k)$. This error is propagated through the boundary location to all other variables.



\subsection{Mullins-Sekerka instability}
\label{stefan:instability}
\begin{figure}
	\centering
	\includegraphics[width=.7\textwidth]{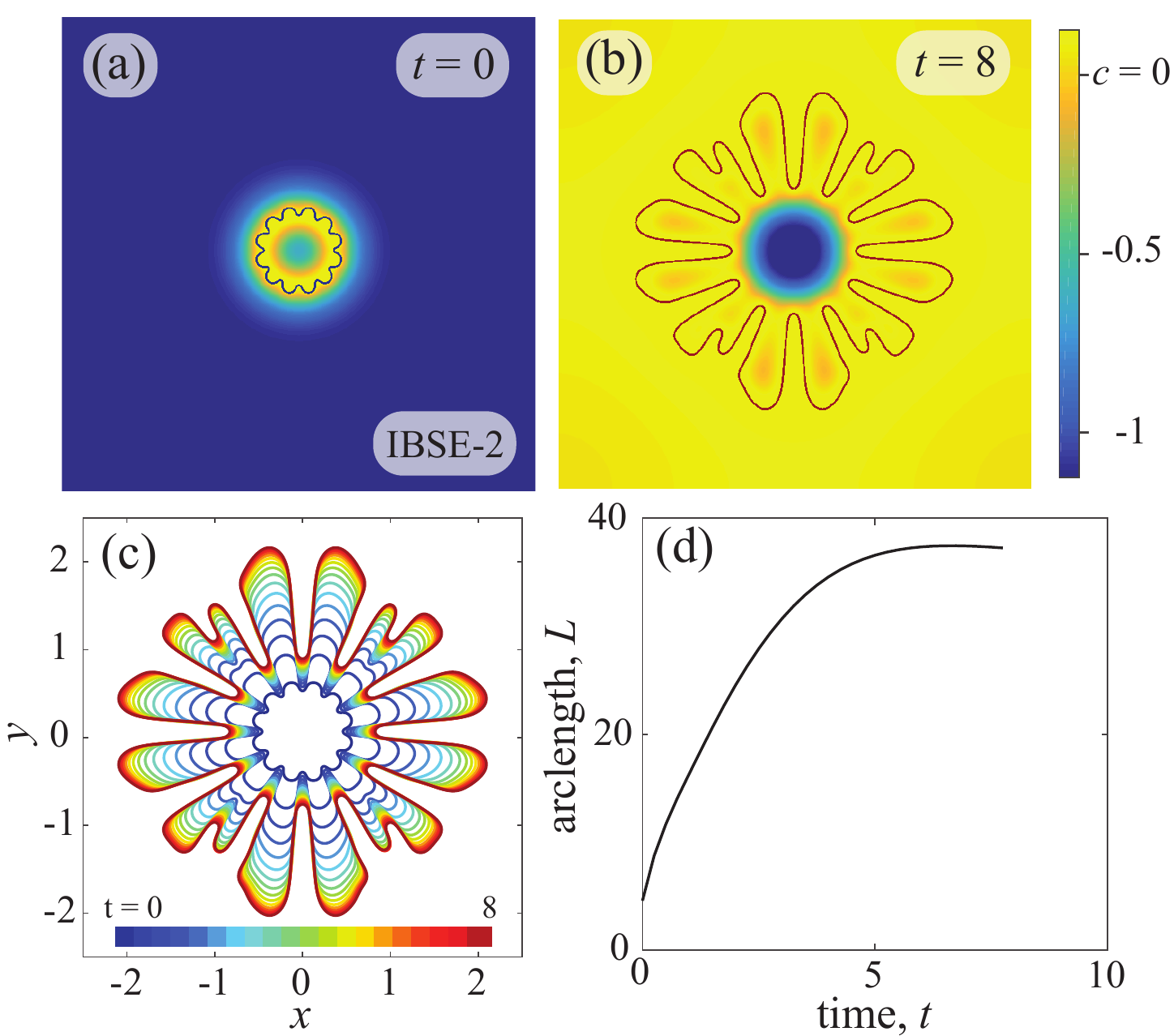}
	\caption{The development of Mullins-Sekerka instability. (a) The initial shape is a circle perturbed with sinusoidal waves. The solution in the physical domain (liquid) and the exterior domain (solid) are both shown. (b) High curvature regions of the interface tend to grow faster, and form finger-like instabilities. (c) Contour plot of the interface. It is clear that the extruding high curvature region grows faster. (d) The arclength $L$ first grows and then saturates. The saturation is due to the temperature reaching equilibrium in the liquid, which is $c = 0$ in the physical domain. Movie of the full simulation can be found in supplemental material S1.}
	\label{fig3}
\end{figure}

The Mullins-Sekerka (MS) instability is frequently observed when solidification occurs in an over-cooled fluid \cite{CHEN19978}. The solid stays at the melting temperature ($c=0$, here), and the temperature gradient determines the rate of liquid solidification. Regions of the interface with higher curvature result in locally higher temperature gradients, enhancing the local solidification, further increasing the local curvature. Thus once the local curvature begins to deviate from the mean curvature, the MS instability leads to growth of those deviations.

In this section, we demonstrate that our solver is able to qualitatively capture the dynamics driven by the MS instability. We start with a perturbed geometry that has $L = \pi$ and $\theta = 0.5 \pi (1-4\alpha) + 0.6\sin{(24\pi \alpha)}$, as shown in \cref{fig3}(a). For this computation, the Stefan number is $\beta = -0.2$ and we use the IBSE-2 method with $N = 300$ and $\Delta t = 5\times10^{-4}$. In order to prevent the formation of singular features on the boundary, regularization via the Gibbs-Thomson effect is required, and we set $\epsilon = 0.002$. \Cref{fig3}(a)-(b) shows the development of the MS instability, and the shape profiles of this process are shown in \cref{fig3}(c). An associated movie is included in supplemental movie S1. The flat part of the boundary grows slowly while the high curvature part grows rapidly, and the resulting growing front further bifurcates into extrusions, which can be seen in \cref{fig3}(c) on the branches at in the NE (Northeast), SE, SW and NW directions. The arclength of the interface is shown in \cref{fig3}(d). Eventually, the growth of the solid is limited by the periodic computational domain and the saturation of temperature in the liquid [\cref{fig3}(b)]. The solidification stops as the liquid reaches an equilibrium temperature of $c = 0$ everywhere. 

\section{Results: Stefan problems coupled with Navier-Stokes flow (dissolution)}
\label{dissolution_results}

\begin{figure}
	\centering
	\includegraphics[width=.4\textwidth]{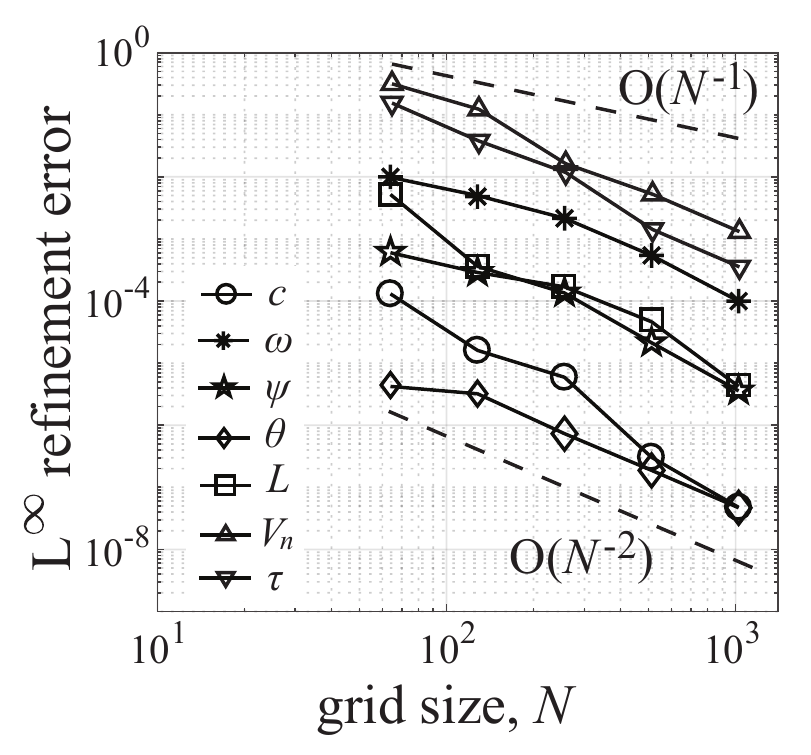}
	\caption{Refinement study of the Stefan solver coupled to an incompressible viscous flow. The spatial discretization and timestep are refined simultaneously so that $\Delta x / \Delta t$ remains constant. The bulk fields $c$ (concentration), $\omega$ (vorticity), and $\psi$ (streamfunction), as well as the surface quantities $\theta$ (tangent angle), $L$ (total arclength), $V_n$ (boundary velocity), and $\tau$ (shear stress) all show second-order convergence in the $L^\infty$ norm (over $\Omega_\textnormal{liquid}$ for bulk fields and $\Gamma$ for surface quantities.)}
	\label{fig4}
\end{figure}

We now turn our attention to the case of a Stefan problem coupled to an incompressible flow. When the Reynolds number is high, the BDF-based IMEX scheme used to timestep pure Stefan problems severely limits the timestep. We instead use a \second-order method based on an Adam-Bashforth Backward-Differentiation (ABBD) scheme. Spatially, we set $k = 2$ in \cref{c-eqn:ibse_poisson,omega-eqn:ibse_stokes}, so $ c\in C^2$, $\omega \in C^1$, $\mathbf{u} \in C^2$ and $\psi \in C^3$.  At time $t$, the set of equations
\begin{subequations}
\label{NS-helmtz-1}
 \begin{align}
 	 \label{NS-helmtz-1-1}
 	 (\mathbb{I} - \sigma_c \Delta) c^t = f_c^t &\quad \mbox{ in  } \Omega_{liquid},\\
 	 \label{NS-helmtz-1-2}
     (\mathbb{I} - \sigma_\omega \Delta) \omega^t = f_\omega^t &\quad \mbox{ in  } \Omega_{liquid},\\
     \Delta \psi^t = - \omega^t &\quad \mbox{ in  } \Omega_{liquid},\\
     \psi = \psi_n = 0, c = 1 &\quad \mbox{ on  } \Gamma,\\
     \psi = \psi_n = 0, \frac{\partial c}{\partial n}=0 &\quad \mbox{ on  } \Gamma_w
\end{align} 
\end{subequations}
is solved, where
\begin{subequations}
\label{ABBD}
 \begin{align}
 	 &\quad\quad\quad\sigma_c = \frac{2\Delta t}{3 \PeN} , \quad \sigma_\omega = \frac{2\Delta t}{3 \ReN}\\
     f_c^t &= -\frac{2\Delta t}{3} \left[2 (\mathbf{u}\cdot\nabla c)^{t-\Delta t} - (\mathbf{u}\cdot\nabla c)^{t-2\Delta t}\right] + \frac{1}{3}\left(4 c^{t-\Delta t} - c^{t-2\Delta t}\right) ,\\
     \label{fomega}
     f_\omega^t &= -\frac{2\Delta t}{3} \left[(\frac{\partial c}{\partial x})^t+2 (\mathbf{u}\cdot\nabla \omega)^{t-\Delta t} - (\mathbf{u}\cdot\nabla \omega)^{t-2\Delta t}\right] + \frac{1}{3}\left(4 \omega^{t-\Delta t} - \omega^{t-2\Delta t}\right).
\end{align} 
\end{subequations}
The concentration \cref{NS-helmtz-1-1} is solved first, so that the value of $\partial c / \partial x$ at time $t$ is known in \cref{fomega}. Anti-aliasing for the non-linear terms is done by smoothly rolling off high-frequency modes, using the method introduced in \cite{hou2007computing}.

\subsection{Convergence of the dissolution solver}

We first ensure that our solver achieves the desired order of accuracy through a refinement study. We let $\ReN = 3.16$, $\PeN= 3.16$, $\beta = 0.1$, and $\epsilon = 0.1$. The aspect ratio of the domain is taken to be $4/3$ (width / height, controlled by adjusting the wall size). We run simulations at grid-sizes of $N=32,64,\dots 512$ with the timestep $\Delta t = 0.01/N$. Each simulation runs for $N$ steps in time so $t_{end} = 0.01$ for all simulations. \Cref{fig4} shows, for both bulk and surface fields, the difference, measured in $L^\infty$, between the solution with a grid size $N$ and $N/2$; achieving the expected second-order convergence. Moreover, our method accurately captures the surface shear stress $\tau = \ReN^{-1} \partial^2\psi /\partial n^2$, enabling us to study near surface flow phenomena such as the boundary layer separation in \cref{highRa}. As with the pure Stefan problem examined in \cref{stefan:convergence}, total accuracy is limited by the estimation of the boundary velocity $V_n$, which is only second-order since it depends on $\partial c/\partial n$ and $c\in C^2(C)$.

\subsection{Dissolution problem in the laminar boundary layer regime}
\label{example:no_separation}

\begin{figure}
	\centering
	\includegraphics[width=0.7\textwidth]{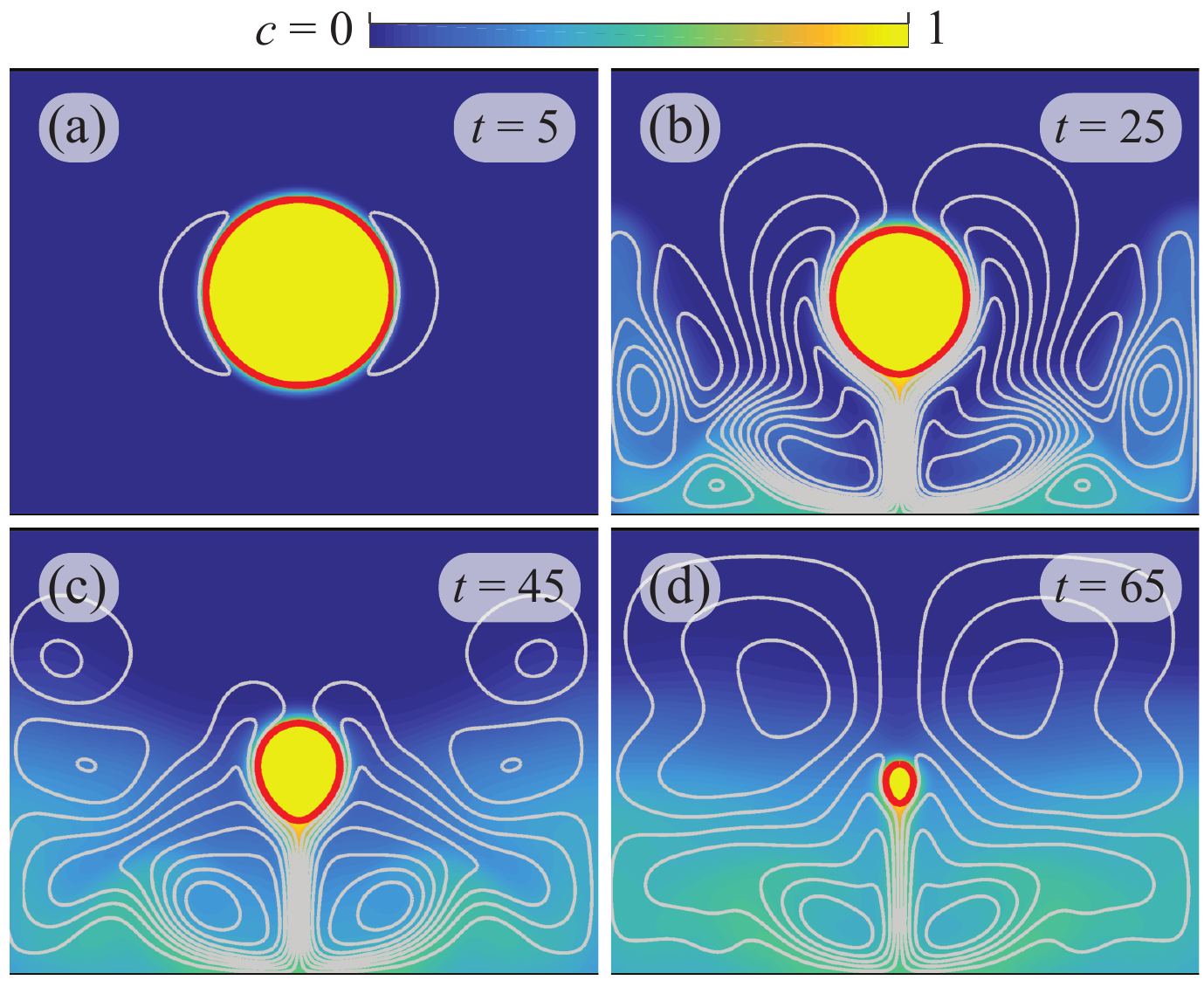}
	\caption{Stefan problem with natural convection. The color-map shows the concentration field $c$, overlaid by contours of the streamfunction $\psi$ . The interface is highlighted in red. A boundary layer forms around the dissolving body, and flow separates at the bottom stagnation point. In this simulation, $\ReN = 316$, $\PeN = 316$ and $\RaN = 10^5$. A corresponding movie can be found in supplemental material S2.}
	\label{fig5}
\end{figure}
\begin{figure}
	\centering
	\includegraphics[width=0.7\textwidth]{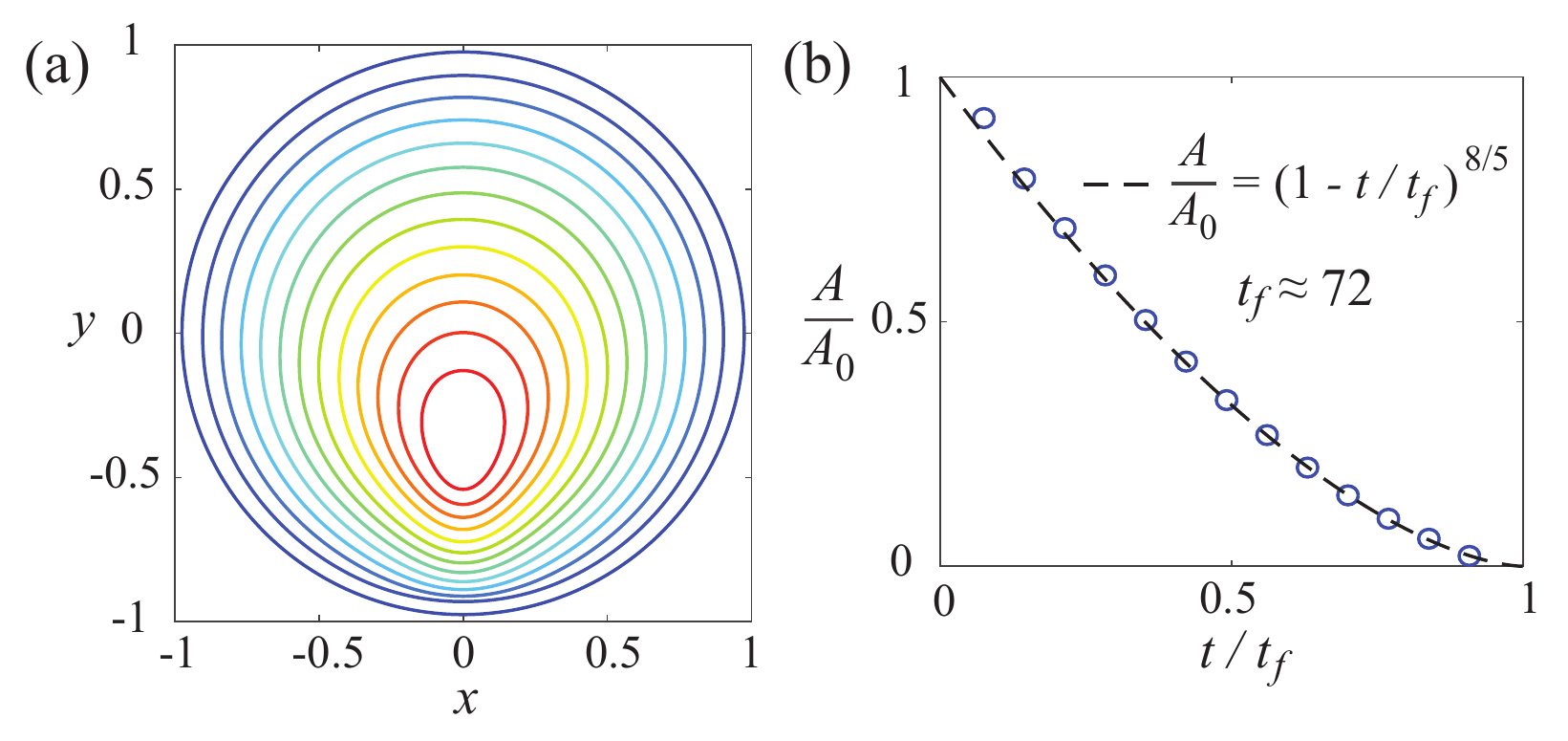}
	\caption{Shape dynamics of the Stefan problem with natural convection shown in \cref{fig5}. (a) The position of the interface as a function of time (early times blue, later times red). At later times, the geometry is asymmetric, and appears egg-shaped. (b) The area of the dissolving body decreases in time with a power law $A/A_0 = (1-t/t_f)^{8/5}$. The numerical value of area (blue circles) matches the predicted power law (dashed line), except for the beginning where the boundary layer is not fully established. }
	\label{fig6}
\end{figure}

We initialize a problem with a circular solid domain immersed in fluid that initially has no solute concentration. When the Reynolds and P\'{e}clet numbers are moderate, the system displays an up-down symmetry breaking, as shown in \cref{fig5,fig6}. In \cref{section:phase_diagram}, we show how the shape dynamics changes across a range of parameters, and argue that pattern formation is driven primarily by increasing the Rayleigh number $\RaN = \PeN\ReN$. For the simulation in \cref{fig5,fig6}, we have set $\ReN = 316$, $\PeN = 316$, $\RaN = 10^5$ and the Stefan number to $\beta/\PeN = 0.003$, while the Gibbs-Thompson effect is set to be low at $\epsilon=5\times 10^{-4}$. The grid size in this simulation is $N = 300$, and the timestep is $\Delta t = 0.004$.

\Cref{fig5} shows the concentration field $c$ and the contour lines of the streamfunction $\psi$. Fluid with higher solute concentration has a higher density than the ambient fluid, and the buoyancy difference results in a gravity driven downward flow. A movie of the simulated dissolution process can be found in the supplement material S2. At the Reynolds and P\'{e}clet number of $316$, a gravity driven flow forms within the boundary layer near the solid, and separates from the body only at the bottom stagnation point. Two vortices are clearly seen below the dissolving body, due to the combination of the downward flow and the splitting flow at the bottom wall. In the later stages of the simulation, a circulation pattern of four vortices can be seen across the computational domain, as shown in \cref{fig5}(d). Overall, the flow is in a laminar regime, and the competition between the viscosity/diffusion and the gravity driven flow leads to the formation of the boundary layer surrounding the dissolving solid.

The time-dependent profile of the interface between the solid and fluid domains is shown in \cref{fig6}(a). The initial trace is shown in blue, and the final trace in red; successive curves are equispaced in time. From the initial circular configuration, the up-down asymmetry develops into an egg-shaped geometry. Comparing the spacing between each interface, we see that the top of the solid domain dissolves fastest, as the fluid near the top separation point is the freshest (that is, has a low solute concentration). Flows around the interface appear to stay attached until the very bottom of the body, and the variation of $c$ is mostly contained in a laminar boundary layer. This enables us to establish a boundary layer scaling. It is known that the concentration gradient within the boundary layer of natural convection has the scaling $\partial c/\partial n \sim L^{-1/4}$ \cite{schlichting2016boundary}, where $L(t)$ is a typical length-scale that we choose as the total arclength. So the boundary velocity $V_n\sim \partial c/\partial n \sim L^{-1/4}$, and the rate of area change can be estimated as 
\begin{equation}
\frac{dA}{dt}\sim LV_n\sim L^{3/4}\sim A^{3/8}.
\end{equation}
Integration yields
\begin{equation}
\frac{A}{A_0} = \left(1-\frac{t}{t_f}\right)^{8/5},
\end{equation}
where $A_0 = A(0)$ is the initial area while $t_f$ is the time at which the solid body vanishes. After fitting this parameter ($t_f \approx 72$), the numerical value of the solid area fits the predicted scaling law, see \cref{fig6}(b). We note that the simulation deviates from the power-law prediction near the beginning; this is to be expected as the boundary layer is not fully established.

Within the laminar boundary layer regime, the interface remains smooth and rounded. As we shall see in the proceeding examples, the boundary layer becomes unstable when the Reynolds and P\'{e}clet numbers are increased, and the interface geometry becomes more complex.

\subsection{Shape dynamics as Rayleigh number increases}
\label{section:phase_diagram}

\begin{figure}
	\centering
	\includegraphics[width=\textwidth]{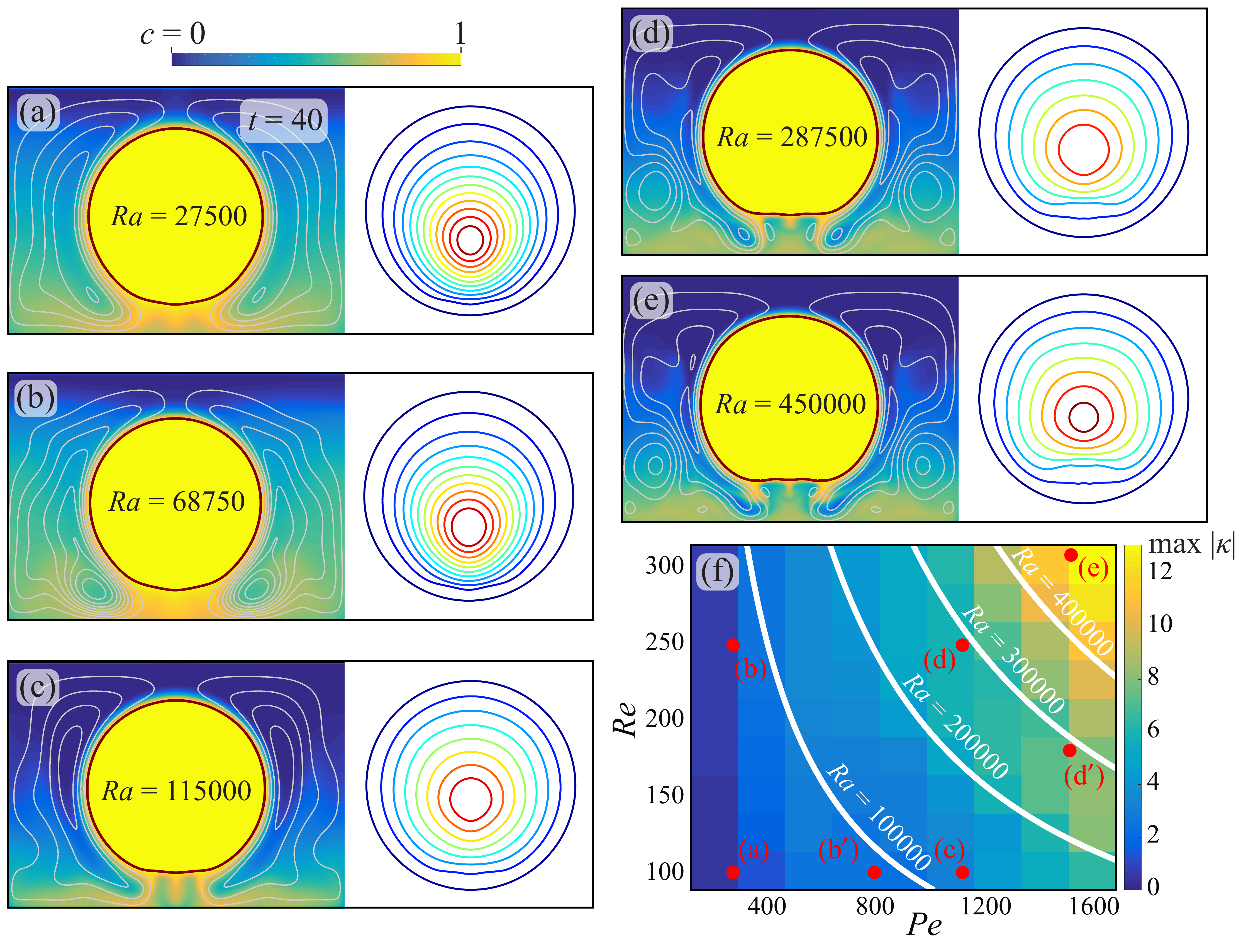}
	\caption{Dissolution at various Rayleigh numbers. (a)-(e) At $t = 40$, with the Stefan number fixed at $\beta / \PeN = 0.003$, higher Rayleigh number results in finer structures in both the concentration field and the solid geometry. Adjacent contours of the interface shape are separated by $\Delta t = 30$. (f) Phase diagram of the maximum curvature of the evolving interface as a function of $\PeN$ and $\ReN$.  Lines of constant Rayleigh number are shown in white, and configurations corresponding to (a)-(e) are labelled in red dots. Movies of (a)-(e) can be found in supplemental material S3-S7.}
	\label{fig7}
\end{figure}

To understand the parameters affecting shape dynamics, we run simulations across a range of Reynolds and P\'eclet numbers. For this parameter sweep, we hold the Stefan number $\beta/\PeN = 0.003$ fixed so that changes in shape dynamics are driven solely by changes in $\PeN$ and $\ReN$. Sweeping through $\PeN$ and $\ReN$ changes the Rayleigh number $\RaN = \ReN\PeN$. \Cref{fig7}(a)-(e) shows 5 simulations with their concentration and flow fields at $t=40$ and their shape evolution in color contours, with parameters: (a) $\ReN = 100$, $\PeN = 275$, $\RaN = 27500$; (b) $\ReN = 250$, $\PeN = 275$, $\RaN = 68750$; (c) $\ReN = 100$, $\PeN = 1150$, $\RaN = 115000$; (d) $\ReN = 250$, $\PeN = 1150$, $\RaN = 287500$; (e) $\ReN = 300$, $\PeN = 1500$, $\RaN = 450000$. Movies corresponding to \cref{fig7}(a)-(e) can be found in the supplemental movies S3-S7. Two observations can be made as the Rayleigh number increases: first, the boundary layer separates before reaching the bottom stagnation point; second, the length-scales present in the concentration, flow fields, and interface become finer.

As a measure of the scale over which patterns are formed, we examine the non-dimensional curvature $\kappa = -\partial \theta / \partial \alpha$.  The actual planar curvature is $\kappa^* = \kappa / L(t)$, which diverges as $L(t)\to0$. In \cref{fig7}(f) we show $\max_{\alpha,t}|\kappa(\alpha,t)|$ for each set of parameters. In all simulations this curvature increases and peaks soon after the moment of boundary layer detachment, and decreases thereafter. The boundary layer detachment is associated with the formation of near-corner regions with high curvature, while the Gibbs-Thomson effect limits the growth of this curvature and diffuses the distribution of $\theta$ over time \cite{moore2013self,Ristroph19606}.

It is apparent that the curvature increases as $\ReN$ and $\PeN$ increase, although neither parameter explains the growth on its own. In the phase diagram \cref{fig7}(f), we plot contours of constant $\RaN$ as white lines. As the maximum curvature is nearly constant on these contours, and increases with increasing $\RaN$, it is apparent that $\RaN$ plays a central factor in determining the pattern morphology.

\begin{figure}
	\centering
	\includegraphics[width=.9\textwidth]{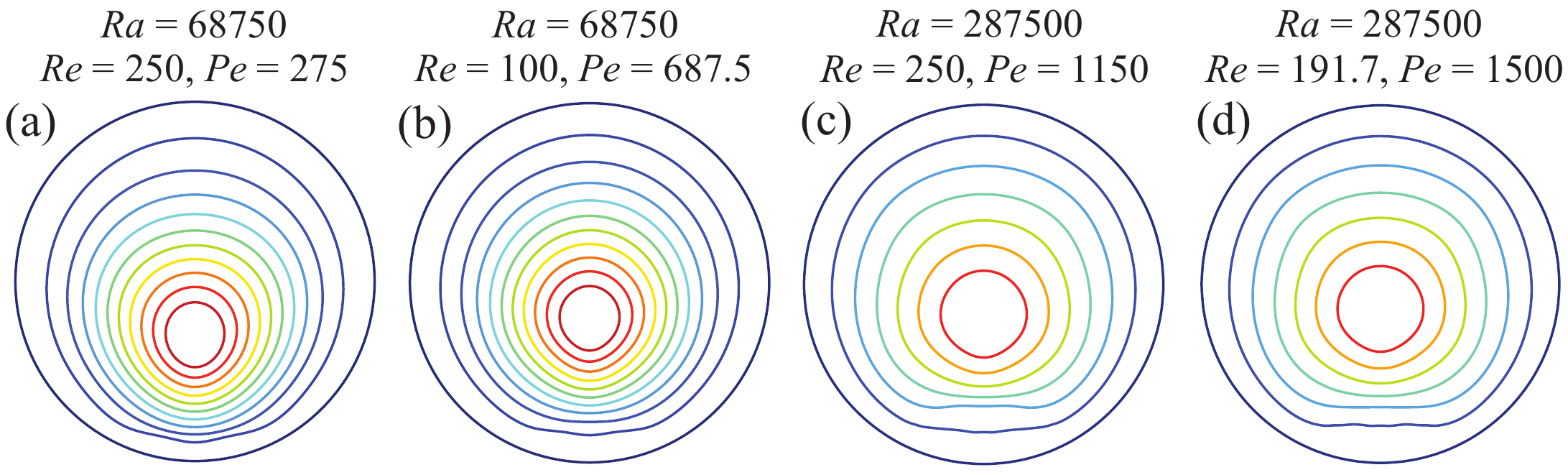}
	\caption{Comparing simulations with the same Rayleigh number but different Reynolds and P\'eclet numbers. (a)-(b) Shape dynamics of dissolution at a low Rayleigh number $\RaN = 6.785\times 10^4$, parameters $\ReN$ and $\PeN$ for (a) and (b) correspond to point (b) and (b') on \cref{fig7}(f). The contours with same color are selected such that they have the same arclength $L$. (c)-(d) Shape dynamics of dissolution at a high Rayleigh number $\RaN = 2.875\times 10^5$, parameters $\ReN$ and $\PeN$ for (c) and (d) correspond to point (d) and (d') on \cref{fig7}(f). The contours with the same color have the same arclength.}
	\label{fig8}
\end{figure}

To further investigate the dependence between dissolving shape dynamics and Rayleigh number $\RaN$, we compare simulations with the same $\RaN$ but different $\ReN$ and $\PeN$. In \cref{fig7}(f), two points (b) and (b') have the same Rayleigh number $\RaN = 6.785\times 10^4$; their dissolving shape dynamics are shown in \cref{fig8}(a)-(b). Although the two simulations have different $\ReN$ and $\PeN$, we see that the resulting dynamics show visual resemblance. This similarity is also seen for higher $\RaN$, as shown in \cref{fig8}(c)-(d) which correspond to the points (d) and (d') in \cref{fig7}(f). One possible explanation for the similarity in shape dynamics is that the density plume size is a function of the Rayleigh number \cite{ahlers2009heat}, so similar $\RaN$ results in similar length scales presenting in the flow/concentration fields and the geometry.

\subsection{Solver behavior as flow structure changes}
\label{section:gmres}

\begin{figure}
	\centering
	\includegraphics[width=0.5\textwidth]{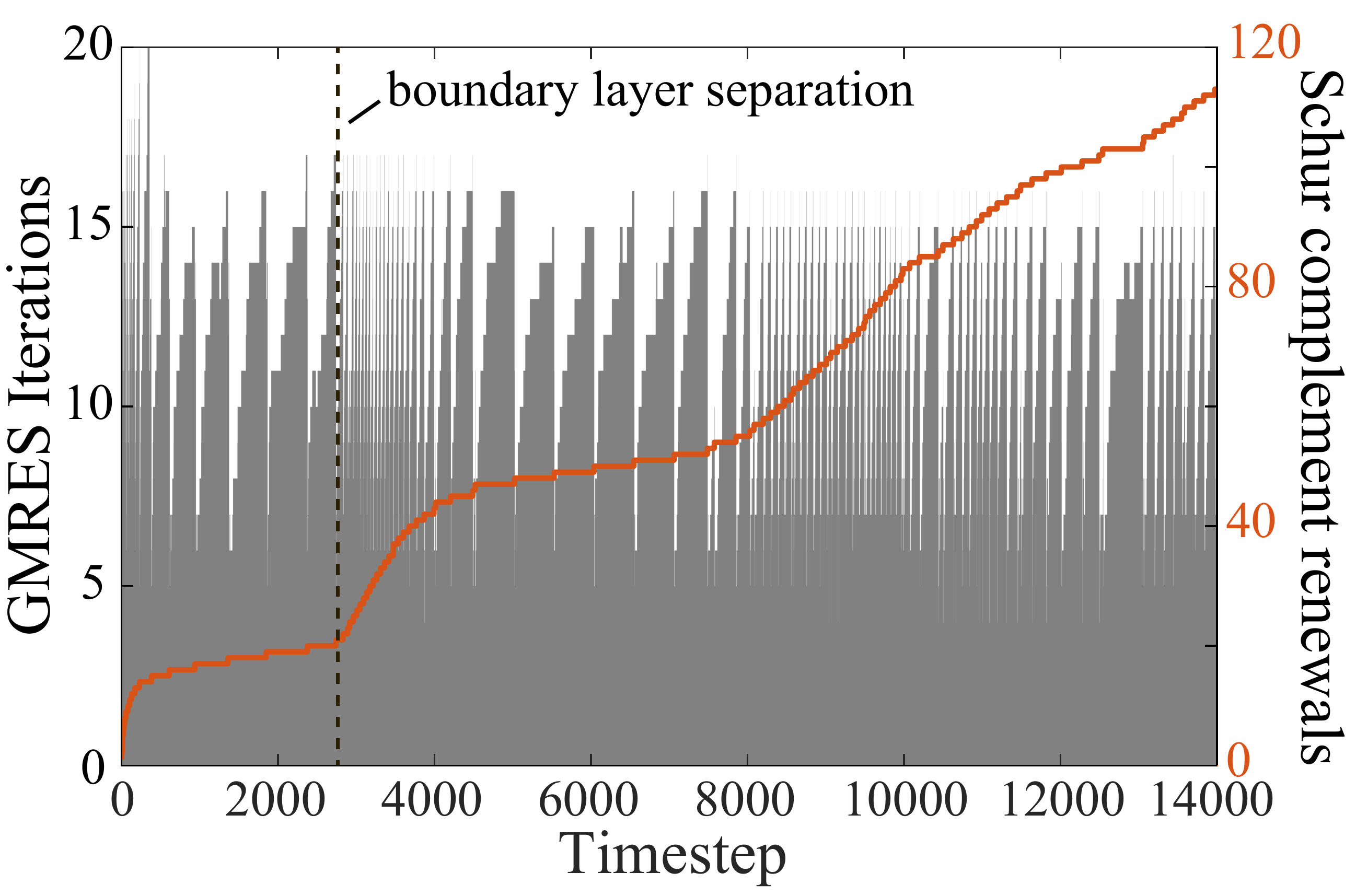}
	\caption{Number of GMRES iterations and renewals of the SC/GSC during a simulation. The Schur complement renews more frequently when the boundary moves rapidly, especially at the beginning of dissolution and at the moment boundary layer separates. Corresponding simulation is shown in \cref{fig7}(d) and supplemental movie S6.}
	\label{fig9}
\end{figure}

SC/GSC renewals are significantly more time consuming than timesteps (although the exact ratio depends on many factors, including the discretization, machine core-count, and the number of GMRES iterations required). To assess the performance of our numerical solver, we show the number of GMRES iterations per timestep and the number of SC/GSC renewals for the simulation shown in \cref{fig7}(d). As shown in \cref{fig9}, the SC and GSC are, on average, renewed every 100 timesteps, resulting in significant amortization of the renewal cost.

At the beginning of the simulation, the initial concentration field is $0$ in the fluid and $1$ at the boundary. The sharp gradient at the dissolving boundary drives rapid boundary motion and the SC renews at a fast pace. As the concentration builds up around the boundary, the renewals become less frequent. After a SC/GSC renewal, the number of GMRES iterations steadily increases as the boundary shape deviates more and more from the shape at last renewal. Rapid boundary motion also happens at the moment of boundary layer separation, where the built up concentration field around the boundary becomes gravitationally unstable and a downward jet forms as the concentration blob drips (supplemental movie S6). As the thickness of the boundary layer $\delta$ suddenly decreases, the resultant concentration gradient $\partial c/\partial n\sim \delta^{-1}$ increases and SC/GSC renews more rapidly due to the boundary motion.

\subsection{Dissolution problem at higher Rayleigh number}
\label{highRa}

\begin{figure}
	\centering
	\includegraphics[width=0.9\textwidth]{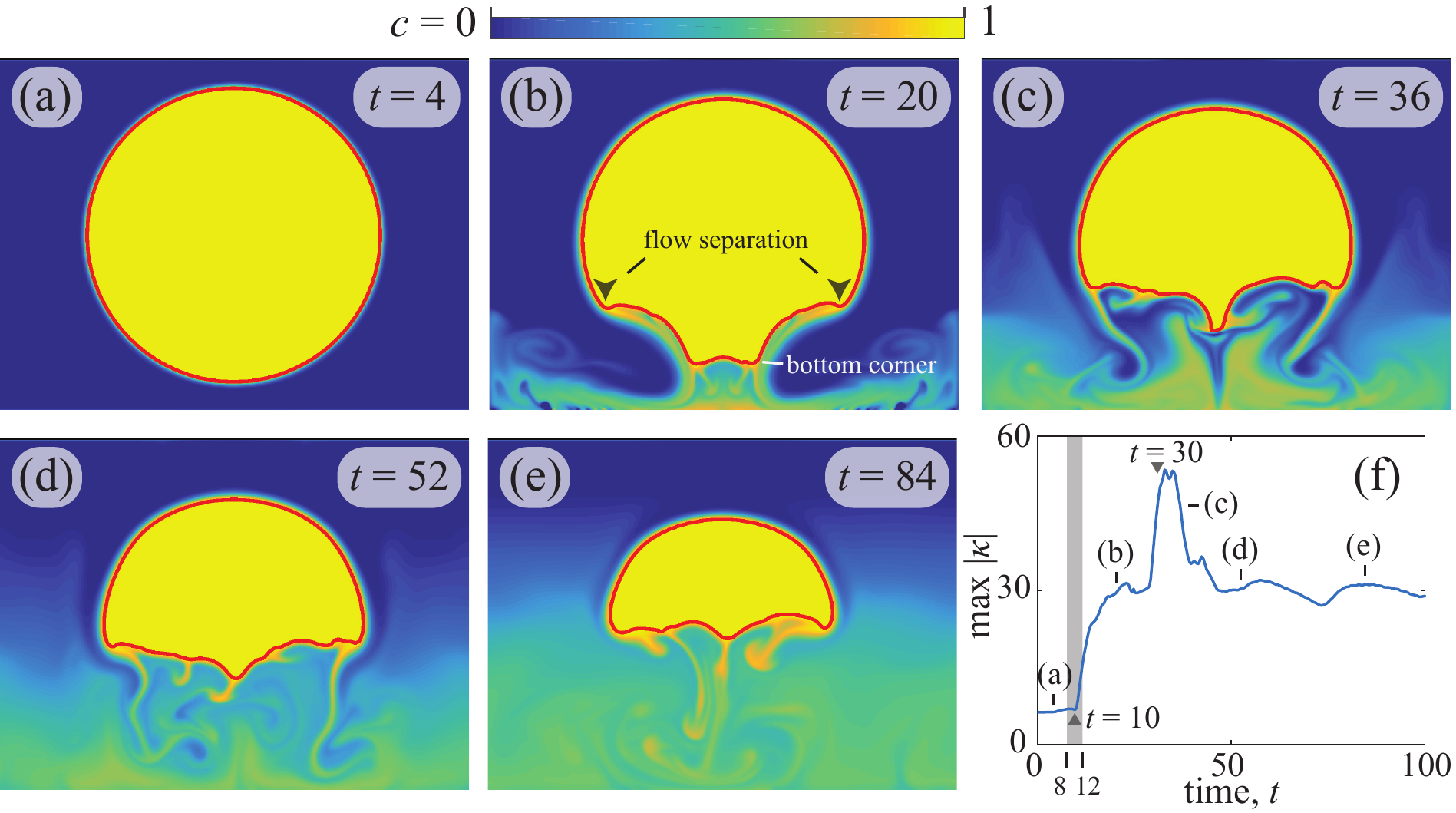}
	\caption{High Rayleigh number of $\RaN = 10^6$ leads to flow separation and pattern formations. (a)-(e) Snapshots of the concentration field and solid geometry at various time. The bottom of dissolving solid shows more complex geometry compared to its top, where dense plumes carrying high concentration fluid detach and sink to the bottom. Movie can be found in supplemental material S8. (f) Maximum non-dimensional curvature $\max |\kappa| = \max_\alpha |\partial\theta / \partial\alpha|$ of the interface grows rapidly as the boundary layer separation occurs at $t = 10$, which signals complex geometry forms and the interface deviates from the circular profile. The bottom corners shown in (b) merge into a single spike at $t=30$, resulting in the maximum curvature over time. }
	\label{fig10}
\end{figure}

In this section we discuss a simulation with the highest Rayleigh number we have achieved so far -- $\RaN = 10^6$ with $\ReN = 316, \PeN = 3160$, $\epsilon = 5\times 10^{-4}$ and Stefan number $\beta / \PeN = 0.005$. \Cref{fig10} shows simulation results using $N = 300$, with $\Delta t = 0.004$. For this Rayleigh number, the flow is no longer laminar [see \cref{fig10}(b) and supplemental movie S8]. Detached density plumes with higher local concentration sink towards the bottom, showing a close resemblance to the thermal plumes of Rayleigh-B\'{e}nard convection \cite{ahlers2009heat, zhang1997non} and the density plumes of the Rayleigh-Taylor instability \cite{tritton2012physical}. Naturally, this change of flow regimes transforms the shape dynamics.

Starting from a circular interface as shown in \cref{fig10}(a), one distinct feature of the dissolution at high $\RaN$ is the formation of near-corners and a rough bottom surface. Once formed, this roughness persists throughout the dissolution process as shown in \cref{fig10}(b)-(e). The maximum non-dimensional curvature $\max_\alpha |\kappa(\alpha,t)|$ is plotted as a function of time in \cref{fig10}(f). Initially, the interface stays very nearly circular with $\kappa \approx 2\pi$. However, the curvature suddenly increases at $t\approx10$ when the boundary layer starts to separate, and the surface roughens thereafter (see supplemental movie S8). The curvature peaks at around $t = 30$ in \cref{fig10}(f), and an inspection of the supplemental movie S8 shows that the two bottom near-corners shown in \cref{fig10}(b) merge into a single downward spike in \cref{fig10}(c)-(e) during that time. The curvature of this downward spike is highest along the interface, and its value decreases in time due to the geometric dissipation from the Gibbs-Thomson effect.

\begin{figure}
	\centering
	\includegraphics[width=0.9\textwidth]{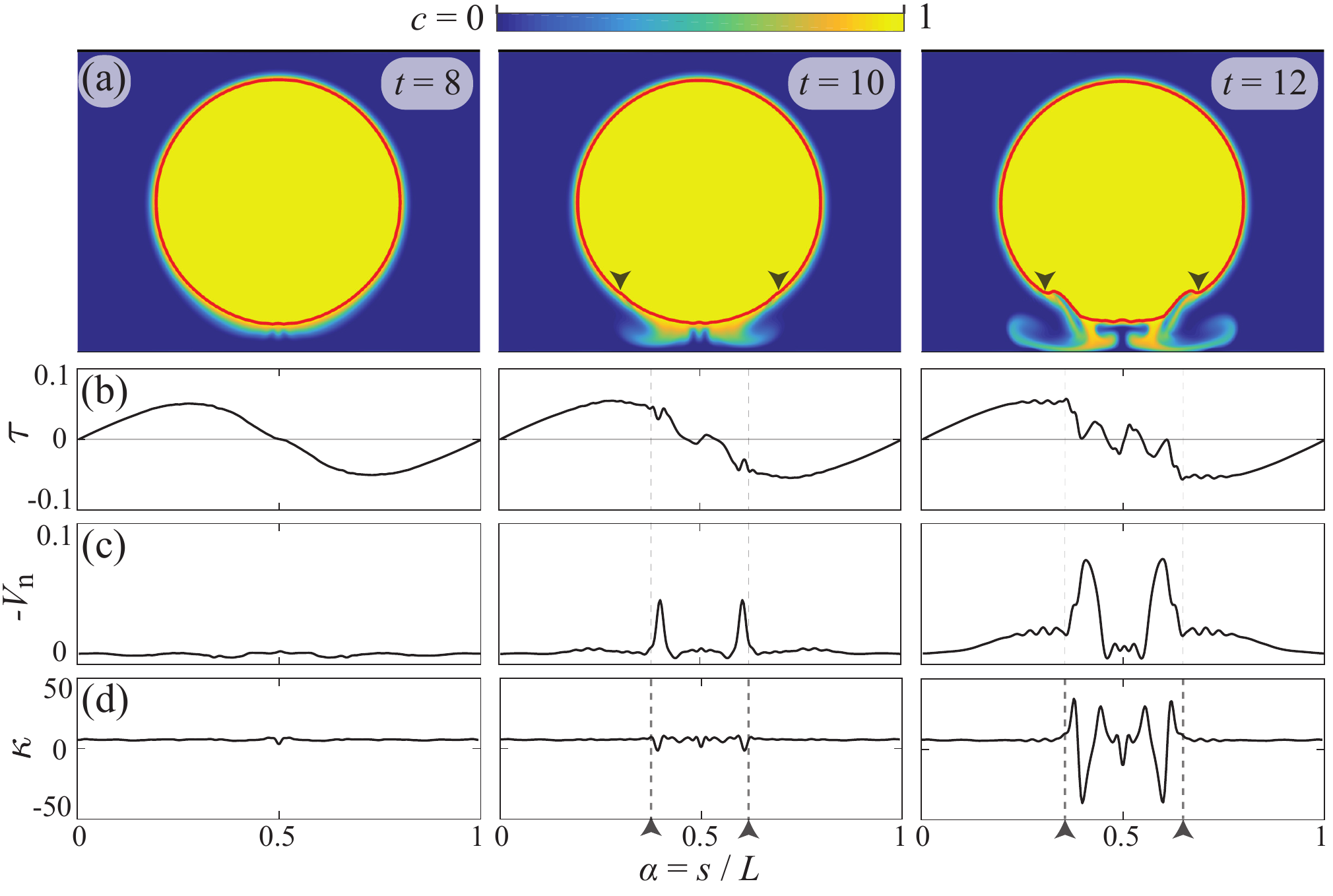}
	\caption{Evolution of the interface and dynamical quantities around the moment of boundary layer separation in \cref{fig10}. The time period of $t=8$-$12$ corresponds to the shaded area in \cref{fig10}(f).  (a) Snapshots of the concentration field around the flow separation at $t = 10$, the separation points are marked with arrows. High concentration plumes form due to the advection of detaching flows. (b) The shear stress $\tau = \ReN^{-1} \partial^2\psi / \partial n^2$ at the moving interface decreases rapidly near the flow separation (marked with dashed lines). (c) The magnitude of normal boundary velocity $-V_n = - \beta \PeN^{-1} \partial c / \partial n$ increases after flow separation. (d) The non-dimensional curvature $\kappa = -\partial_\alpha\theta$ deviates from its initial profile of $\kappa =  2\pi$ as $V_n$ drives the boundary away from a circle. }
	\label{fig11}
\end{figure}

What initiates this formation of fine-scale structure? \Cref{fig11}(a) shows several snapshots of the interface and the concentration field around the time of the first boundary layer separation. The boundary layer has thickened by $t\approx8$, and by $t\approx10$, the bottom boundary layer has detached, eventually forming two heavy plumes (by $t\approx12$) that sink towards the bottom. During the same period of time, near-corners start to develop at the location of flow separation; similar corner formation has been observed in the experiments on erosion \cite{Ristroph19606} and dissolution \cite{FLM:9533822}. One sign of a destabilizing boundary layer is the weakening of surface shear stress $\tau = \ReN^{-1} \partial^2\psi / \partial n^2$ \cite{schlichting2016boundary}. \Cref{fig11}(b) shows such a weakening between the locations of flow separation, marked with dashed lines, between $t=10$ and $t=12$. Indeed, $\tau$ decreases markedly in magnitude there and then. The dissolution rate increases significantly as the flow separates (see \cref{fig11}(c)), due to recirculating flows in the wake that introduce relatively fresh liquid, creating a higher local concentration gradient at the surface.  Likewise, the curvature distribution in \cref{fig11}(d) deviates from its initial profile of $\kappa = 2\pi$, as the bottom of the dissolving solid forms new near-corners. By $t\approx 12$ there are now two descending plumes on each side, one emanating from a newly formed near-corner, and the other from the initial boundary layer detachment point. As the bottom surface of the solid recedes upwards, at later times these plumes destabilize to transverse bending, and the flows become much more complex, as seen in \cref{fig10}.

Small numerical artifacts begin to appear in this simulation at $t\approx12$ (see \cref{fig11}(b)-(d)). In this simulation, the high $\PeN$ and $\ReN$ produce fine length-scales and fast velocities, especially when boundary-layer separation occurs. To compute this solution in a reasonable timeframe, we are using a discretization that just resolves the spatial scales and timesteps that near the Courant-Friedrichs-Lewy condition. Increased temporal and spatial resolution will be required to attack problems with higher $\ReN$ and $\PeN$.  We discuss some possibilities for how to achieve this in \cref{discussion}.

\section{Discussion}
\label{discussion}

In this paper we introduce and test an accurate, sharp-interface method for solving the Stefan problem coupled to a Navier-Stokes flow (the dissolution problem). These problems are numerically challenging, as the boundary motion depends on the normal derivative of field variables. The solver introduced here is based on the IBSE method, which allows accurate resolution of these normal derivatives \cite{Stein2016252}. In refinement studies the method achieves third-order accuracy for Stefan problems and second-order accuracy for the dissolution problem, measured in $L^\infty$ for all variables. In addition, we explicitly verify that the method provides convergent estimates for the normal-derivative of the concentration and the surface shear stress \emph{on the boundary}. Having access to accurate estimates of these quantities allows for an analysis of the way flow structures affect the dissolution process, as in our discussion in \Cref{highRa}. For the classical Stefan problem (without flow), we validate the solver against a known analytic solution and show that it is able to reproduce the well-known Mullins-Sekerka instability. For the dissolution problem, when boundary layer separation does not occur, the solver reproduces a predicted boundary layer scaling, and when boundary layer separation does occur, it qualitatively reproduces interface morphologies observed in experiments \cite{wykes2018self}.

\begin{figure}
	\centering
	\includegraphics[width=0.6\textwidth]{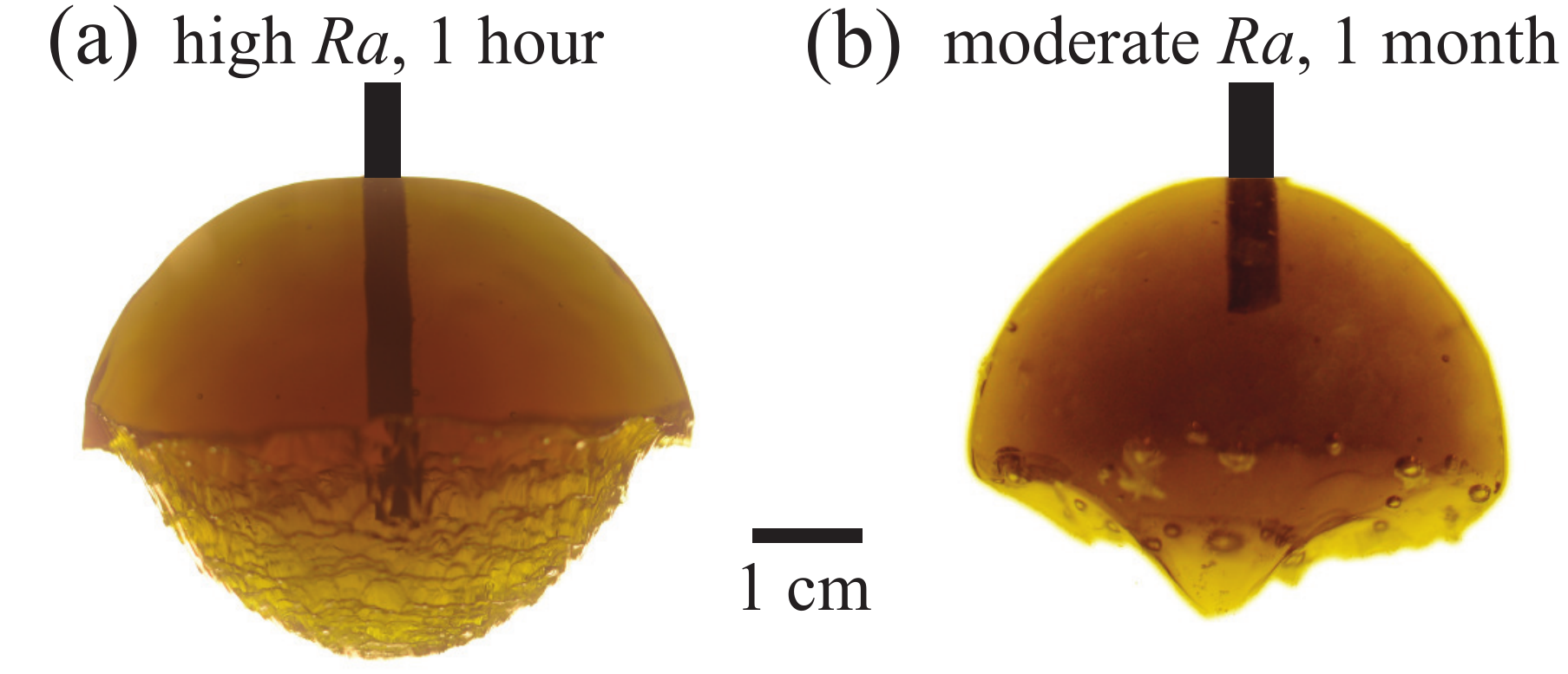}
	\caption{Two experiments of spherical solids of sugar dissolving in water. (a) When the sugar dissolves into fresh water, complex surface textures form at the bottom of the dissolving object \cite{wykes2018self}; parameter values are $\ReN \approx 10^4$, $\PeN \approx 10^7$ and $\RaN \approx 10^{11}$. (b) Adding sugar to the background fluid results in a moderate Rayleigh number ($\RaN\approx4\times10^5$), and surface textures with a larger length-scale are observed. In this experiment a prominent downward tip forms, located at the bottom of the dissolving solid. The shape of the dissolving solid is qualitatively similar to the shape observed in the numerical simulation shown in \cref{fig10}; the Rayleigh numbers are similar despite radically different Reynolds and P\'eclet numbers.}
	\label{fig12}
\end{figure}

Some experimental observations of dissolution are found in \cref{fig12}, which shows an  initially spherical solid made of sugar dissolving into a large body of liquid. \Cref{fig12}(a) shows the dissolution of sugar into fresh water. The dissolved shapes have scalloped patterns on the bottom and a sharp edge separating the rough bottom from its smooth top, where the flow has a laminar boundary layer structure. In \Cref{fig7}, we observed from our simulations that a higher Rayleigh number is associated with finer structures in the flow/concentration fields and higher curvatures in the surface morphology, so the formation of fine patterns in \cref{fig12}(a) is expected, although we were unable to compute numerical solutions in this parameter regime, where $\ReN \sim 10^4$, $\PeN \sim 10^7$, and $\RaN\sim10^{11}$. Similar patterns have been observed experimentally in dissolution \cite{wykes2018self,claudin_duran_andreotti_2017} and numerically in melting \cite{favier_purseed_duchemin_2019}.

In the experiment shown in \cref{fig12}(b), sugar has been added to the initial fluid into which the solid dissolves, increasing the viscosity and decreasing the Reynolds number of the associated flow. Since the normal velocity $V_n \propto \beta/\PeN$ and we know the P\'eclet number for the experiment shown in \cref{fig12}(a) \cite{wykes2018self}, we can compare the timescales of dissolution between the two experiments shown in \cref{fig12} to estimate the P\'eclet number associated with \cref{fig12}(b) to be $\PeN\approx10^{10}$. To estimate the Rayleigh number for this experiment, we use the Stokes-Einstein relation \cite{miller1924stokes} to obtain $\nu D = k_B T/(6\pi R \rho) \approx 10^{-16}\mbox{ m}^4/\mbox{s}^2$, where $k_B$ is the Boltzmann constant, $T \approx 300$ K is the temperature, and $R \approx 1$ nm is the radius of a sugar molecule. Using the shadowgraph technique\footnote{\url{https://en.wikipedia.org/wiki/Shadowgraph}}, we observe that the density plumes move approximately 1 cm every minute so that the typical flow speed is $U\approx 10^{-4}$ m/s. With a typical length-scale of $L\approx 6$ cm which is the initial diameter of the sphere, the Rayleigh number in \cref{fig12}(b) is $\RaN = \ReN\PeN = (UL)^2 / (\nu D) \approx 4\times10^5$ -- a number that is very close to the Rayleigh number from the simulation shown in \cref{fig10}. In \cref{fig8}, we observed that simulations with the same Rayleigh number but different Reynolds and P\'eclet numbers had similar morphologies. Remarkably, the dissolved shapes in \cref{fig10} and \cref{fig12}(b) show a qualitative resemblance even though the parameters in the experiment ($\PeN\approx 10^{10}$ and $\ReN = \RaN/\PeN \approx 10^{-4}$) are drastically different from those used for the simulation ($\PeN = 3160$ and $\ReN = 316$). The relationship between the interface morphology and the Rayleigh number suggests that the interaction between the flow field, concentration field, and the geometry is critical in driving these complex shape dynamics. The numerical method developed in this paper allows us to accurately compute important quantities that are hard to measure experimentally, including the surface shear stress and concentration gradients near to the boundary. This extra information is crucial in helping piece together a full understanding of the shape dynamics of this system.

In the experiments shown in \cref{fig12}(a) the Rayleigh number is about $10^{11}$, with a Schmidt number of $10^3$; corresponding to a Reynolds number $\ReN\sim10^4$ and a P\'{e}clet number of $\PeN \sim 10^7$. Unfortunately, these parameter values lead to extremely thin boundary layers (in both $\mathbf{u}$ and $c$), requiring high resolution to accurately resolve. In fact, most small solute molecules yield a Schmidt number around $\ScN = \nu/D = \PeN/\ReN \sim 10^3$ in water due to their similar molecular diffusivity at $D\sim10^{-9}$ m$^2/$s \cite{incropera2007fundamentals}, which leads to unavoidably fine scales in studying dissolution problems at moderate or high $\ReN$. At low $\ReN$, the Stokes-Einstein relation enforces $\PeN\sim\ReN^{-1}$, which adds stiffness to the advection-diffusion equation and again leads to thin boundary layers. Numerically studying these problems in experimentally relevant parameter regimes will require the use of an adaptive solver to capture these fine near-boundary flow structures. The Immersed Boundary Method, on which the IBSE method is based, has been implemented in an adaptive framework in IBAMR\footnote{\url{https://ibamr.github.io/}}, and extending the methods developed in this paper within that framework presents one possible method for attacking such problems. Nevertheless, there are other interesting problems that are accessible to the solver as developed, which can, on a standard workstation, handle Reynolds and P\'{e}clet number numbers of $\ReN \sim 300$ and $\PeN \sim 3000$, respectively. These are approximately the parameter values for ice melting into water \cite{favier_purseed_duchemin_2019}. This situation also demands special care, as water has a density anomaly near $T = 4^\circ$C; thus, the Boussinesq approximation used in this paper must be modified to correctly represent the known buoyancy-temperature dependence.


\section*{Acknowledgement}
We thank Scott Weady and Leif Ristroph for useful discussions and the National Science Foundation for support (NSF CBET-1805506).
\bibliographystyle{unsrtnat}
\bibliography{refs,refs2}

\end{document}